\documentclass{article}

\usepackage{amsmath,amsfonts,bm}

\def\eqref#1{equation~\ref{#1}}

\def\1{\bm{1}}

\DeclareMathAlphabet{\mathsfit}{\encodingdefault}{\sfdefault}{m}{sl}
\SetMathAlphabet{\mathsfit}{bold}{\encodingdefault}{\sfdefault}{bx}{n}

\usepackage{multirow}
\usepackage{hyperref}
\usepackage{url}
\usepackage{booktabs}
\usepackage{adjustbox}
\usepackage{colortbl}
\usepackage[table]{xcolor} 
\usepackage{caption}
\usepackage{graphicx}
\usepackage{subcaption}
\usepackage{amsmath,amssymb} 
\usepackage{placeins}
\usepackage{graphicx}
\usepackage{tabularx} 
\usepackage{array}
\usepackage{tikz}
\usepackage{amsmath,amssymb,amsthm}

\newtheorem{proposition}{Proposition}

\usepackage{amsmath} 
\usepackage{microtype}
\usepackage{graphicx}
\usepackage{booktabs} 

\usepackage{hyperref}

\usepackage[accepted]{mlsys2025_arxiv}

\mlsystitlerunning{FREESH: Fair, Resource- and Energy-Efficient Scheduling for LLM Serving on Heterogeneous GPUs}

\begin{document}

\twocolumn[
\mlsystitle{FREESH: Fair, Resource- and Energy-Efficient Scheduling for LLM Serving on Heterogeneous GPUs}



\mlsyssetsymbol{equal}{*}

\begin{mlsysauthorlist}
\mlsysauthor{Xuan He}{equal,to}
\mlsysauthor{Zequan Fang}{equal,goo}
\mlsysauthor{Jinzhao Lian}{equal,ed}
\mlsysauthor{Danny H.K. Tsang}{to}
\mlsysauthor{Baosen Zhang}{goo3}
\mlsysauthor{Yize Chen}{goo4}
\end{mlsysauthorlist}

\mlsysaffiliation{to}{Hong Kong University of Science and Technology (Guangzhou), China}
\mlsysaffiliation{goo}{Huazhong University of Science and Technology, China}
\mlsysaffiliation{ed}{Renmin University of China, China}
\mlsysaffiliation{goo3}{University of Washington, USA}
\mlsysaffiliation{goo4}{University of Alberta, Canada}

\mlsyscorrespondingauthor{Yize Chen}{yize.chen@ualberta.ca}

\mlsyskeywords{Machine Learning, MLSys}

\vskip 0.3in

\begin{abstract}
The ever-increasing computation and energy demand for LLM and AI agents call for holistic and efficient optimization of LLM serving systems. In practice, heterogeneous GPU clusters can be deployed in a geographically distributed manner, while LLM load also observes diversity in terms of both query traffic and serving patterns. LLM queries running on advanced GPUs during a high-emission hour at one location can lead to significantly higher carbon footprints versus same queries running on mid-level GPUs at a low-emission time and location. By observing LLM serving requirements and leveraging spatiotemporal computation flexibility, we consider the joint routing and scheduling problem, and propose FREESH to cooperatively run a group of data centers while minimizing user-specified carbon or energy objectives. FREESH identifies the optimal configurations of balanced load serving by matching distinct GPU instance's power-throughput characteristics with predictable LLM query length and workloads. To ensure both latency and fairness requirements, FREESH identifies optimized parallelism and query routing schedules together with dynamic GPU frequency scaling for power saving, and Least-Laxity-First (LLF) serving strategy for query scheduling. During the 1-hour serving on production workloads, FREESH reduces energy by $28.6\%$ and emissions by $45.45\%$ together with improvements in SLO attainment and fairness. 
\end{abstract}
]



\printAffiliationsAndNotice{\mlsysEqualContribution} 

\section{Introduction}

Surging deployment of power-intensive LLM inference workloads across worldwide data centers has raised concerns over AI model's energy and carbon footprint~\cite{maji2025data, lacoste2019quantifying,luccioni2024power}. As the energy needed for inference is set to surpass the already massive quantity of energy used for AI training, inference costs are also increasingly dominated by GPUs' energy consumption~\cite{qiu2024power, samsi2023words}. Cloud service platforms and model providers now deploy a large amount of model replicas across clusters of GPUs to meet huge and stochastic LLM workloads~\cite{zheng2023efficiently, kwon2023efficient}. Although these deployments are homogeneous on the computing side, they can be powered in quite heterogeneous ways. For example, a query served by a GPU instance co-located with solar panels during a sunny day leads to significantly lower carbon emissions compared to those using power from fossil fuel generators~\cite{lisa2025fair}. On top of spatiotemporal differences in energy production, there are significant variations in terms of service-level objectives (SLOs), query lengths, arrival patterns and GPU instance configurations in each LLM serving tasks. Together, they make addressing the energy and emission efficiency of LLM serving a non-trivial problem.


Most serving systems today are still managed using a reservation-based allocation model for GPU provisioning based on peak load. Further, systems either apply heuristics to change GPU frequency~\cite{wu2014green, tang2019impact}, or fix parallelism and scheduling strategies throughout a large time window. Because of inherent stochasticity, these strategies can lead to unbalanced workloads, computationally- and power-inefficient GPU working states, Head-Of-Line (HOL) blocking, and excessive energy consumption. 
Data centers as a whole can serve as assets in the grid~\cite{wierman2014opportunities} by responding to emission and price signals (e.g., demand response) and optimizing operations accordingly~\cite{colangelo2025turning}, although they still operate independently.   Given the unique inference characteristics and stringent request-level constraints such as latency and throughput, jointly optimizing routing and scheduling across multiple GPU clusters has the potential to increase the efficiency and sustainability of large LLM services~\cite{kakolyris2024slo, kakolyris2025throttll}. 





In this work, we tackle the dual problems of emission- and resource-efficient joint LLM routing-scheduling for a group of distributed and heterogeneous GPU servers. This method allows us to take advantage of distributed renewable generation, which could otherwise be curtailed because of temporal and spatial mismatches with computing demand or power transmission limits~\cite{li2024unseen, alizadeh2015dynamic}. In addition, most of current LLM services adopt First-Come-First-Serve (FCFS) scheduling strategy~\cite{kwon2023efficient}. But it can cause HOL blocking and fairness issues~\cite{sheng2024fairness, fu2024efficient}, and is also oblivious to time-varying electricity prices and carbon emission intensities. Therefore, we explicitly account for the heterogeneity in locational emission rate, GPU model's power consumption rate and service rate, locational and temporal patterns of LLM traffic, and propose a cross-layer, multi-timescale solution. We first characterize the interplay between control knobs including LLM request type partition, GPU frequency, Tensor Parallelism (TP), scheduling priority, and LLM serving's performance in terms of SLO, energy/carbon emissions, and fairness. Several recent studies have looked at the tradeoff between LLM inference's energy efficiency and throughput~\cite{stojkovic2025dynamollm, kakolyris2024slo}, and have concluded that lowering TP parallelism and GPU frequency would lead to more energy-efficient operation without significant degradation in performance~\cite{patel2024characterizing}. Our work generalizes this analysis across multiple locations and time horizons and control knobs, and proposes a holistic framework to incorporate both \textit{slow-timescale optimization} on geographically distributed server configurations, and \textit{fast-timescale DVFS adaptation and laxity-based workload scheduling} in response to instantaneous workload traffic. 

In the upper level of FREESH, based on predicted incoming workloads and nodal carbon intensity, a tractable mixed integer program (MILP) problem is formulated to account for emission or energy consumption objectives subject to GPU instance availability and SLOs, which guides model parallelism, load partition, and scaling instance numbers. 
FREESH also achieves load balancing across geographically distributed GPU instances by dynamically partitioning query types based on incoming traffic. At lower level, observing the fast-changing states of arriving requests and inspired by congestion control in networking research~\cite{mo1999analysis, qiu2024power}, we propose an adaptive multiplicative increase, additive decrease (MIAD) frequency scaling scheme to dynamically optimize GPU's power profiles, which achieves satisfiable throughput and minimal power consumption simultaneously. We prove that MIAD converges to the optimal solution in term of efficiency and fairness. Furthermore, utilizing a lightweight query type forecasting model, we can quantify each incoming query's \emph{urgency} level, and propose a novel LLF scheme, which explicitly prioritizes the most urgent query to balance fairness and performance under scaled GPU frequency. Together, this multi-layered architecture collaboratively manages these competing objectives for efficient LLM serving in a flexible manner. Evaluations across diverse emission and production workloads demonstrates that FREESH  achieves a superior trade-off among the competing objectives of cost, SLO, and fairness consistently, especially substantial emission (45.45\%) and energy (28.6\%) reduction are recorded. FREESH is open sourced for evaluation and benchmarking\footnote{\url{https://github.com/AndrewFangZequan/LLM_Serving_FREESH}}.


\section{Background}
\label{headings}

\subsection{Opportunities in LLM Load and Cluster Routing}
    
\textbf{Opportunity 1: Emission reduction from variability in carbon intensity and workload.} Carbon intensity varies both temporally and spatially because of the dynamic nature of electricity grids and their access to renewable energy sources (RES). This indicates that \emph{emissions and energy associated with each token are not distributed equally.} Intuitively, serving large LLM queries from GPU clusters in regions with lower carbon intensity can significantly reduce emissions. This variability allows for the efficient use of low-carbon electricity by shifting computational loads to optimal regions \cite{radovanovic2022carbon, liu2011greening}, which can be achieved through two main strategies: either by activating more servers in low-carbon areas or by routing time-varying query workloads to those regions. Ultimately, a promising approach is to use an optimization model to dynamically select the best locations, while utilizing fine-grained frequency scaling to precisely control GPU power. 
\begin{figure} [ht]
      \centering
      \vspace{-1em}
      \includegraphics[width=0.9\linewidth]{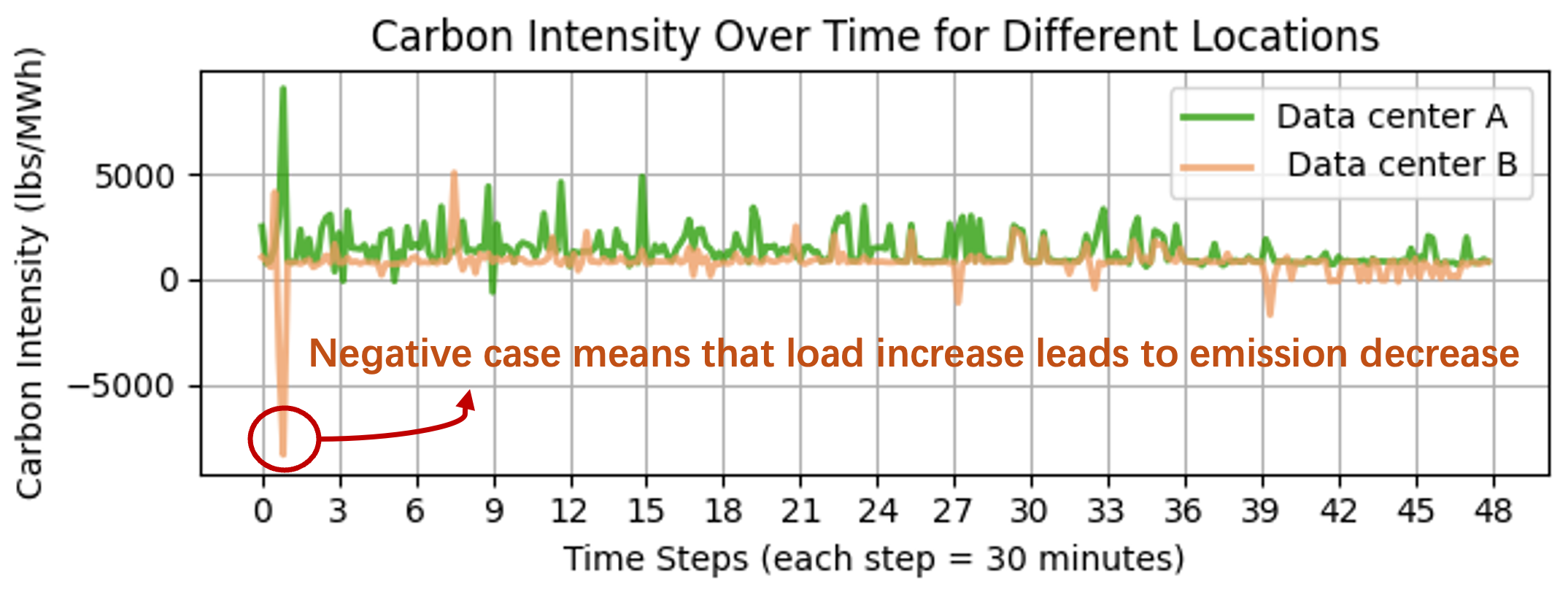} 
      \vspace{-2em}
    \end{figure}
    
\begin{figure} [ht]
      \centering
      \includegraphics[width=0.85\linewidth]{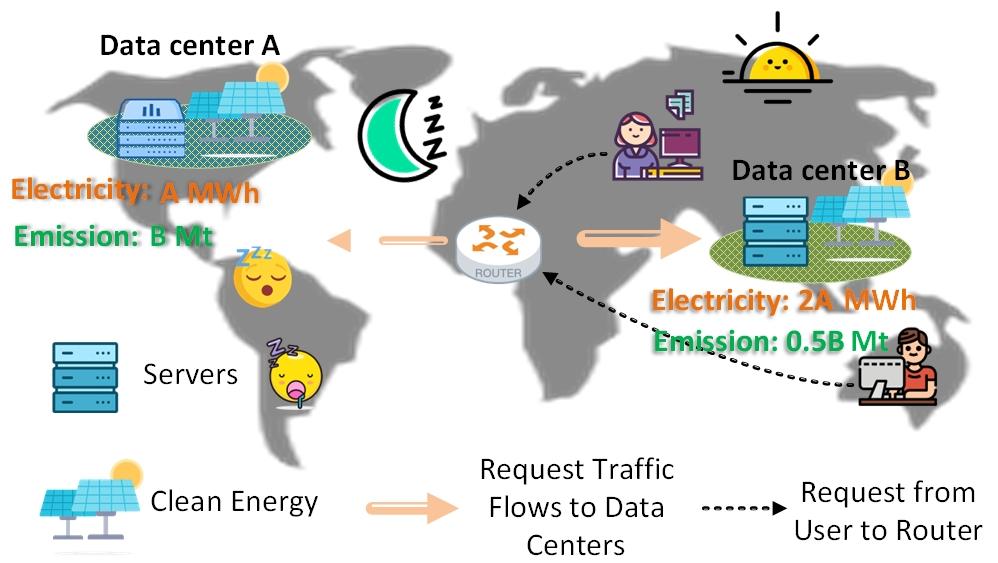} 
      \vspace{-1em}
      \caption{Spatiotemporal routing of LLM load for carbon and energy reduction. Locational marginal emission timeseries for these two data centers are distinctive, giving opportunities for coordinating GPU clusters adaptive to incoming workload characteristics.} \label{fig:Spatiotemporal}
       \vspace{-1em}
    \end{figure}

\begin{figure}[t]
    \centering
    \includegraphics[width=1\linewidth]{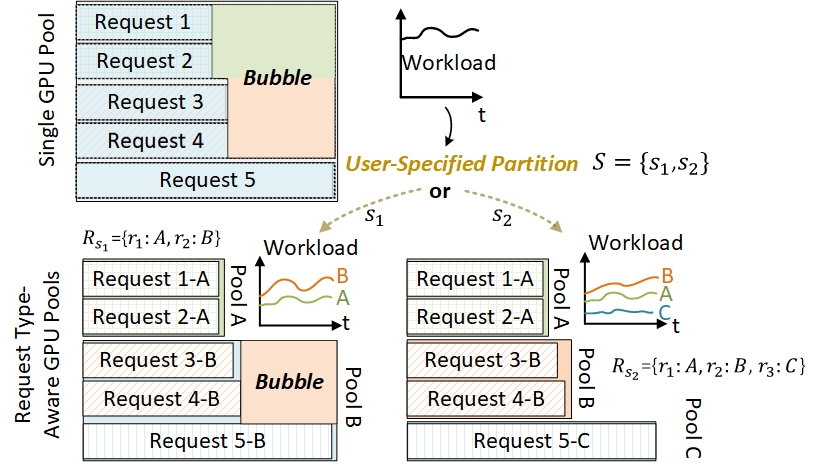}
    \vspace{-2em}
    \caption{Resource utilization improved by request partition.}
    \label{fig:T1_Partition}
\end{figure}

\paragraph{Opportunity 2: Energy saving from the adaptive request partition and work mode configuration.} 
\cite{stojkovic2025dynamollm} proposes to divide clusters into multiple instance pools, where each pool is dedicated to a specific request type. As illustrated in Fig. \ref{fig:T1_Partition}, this design reduces resource allocation "bubbles" compared to the traditional single-pool architecture, where these bubbles indicate the non-occupied or idled allocated GPUs. Notably, the way requests are partitioned based on their input/output lengths directly shapes the system’s workload pattern, especially the throughput of each request type, which in turn dictates the appropriate worker configuration to guarantee the SLO attainment while optimizing the user-specified objective. Compared to static partition adopted in the literature~\cite{stojkovic2025dynamollm}, 
leveraging dynamic request partitioning enables sustained load balancing, a capability that can significantly improve resource utilization and realize the trade-off between multiple serving objectives, but has yet to be explored in existing literature.

\subsection{Opportunities in LLM Scheduling}
\begin{figure}[t]
    \centering
    \includegraphics[width=1\linewidth]{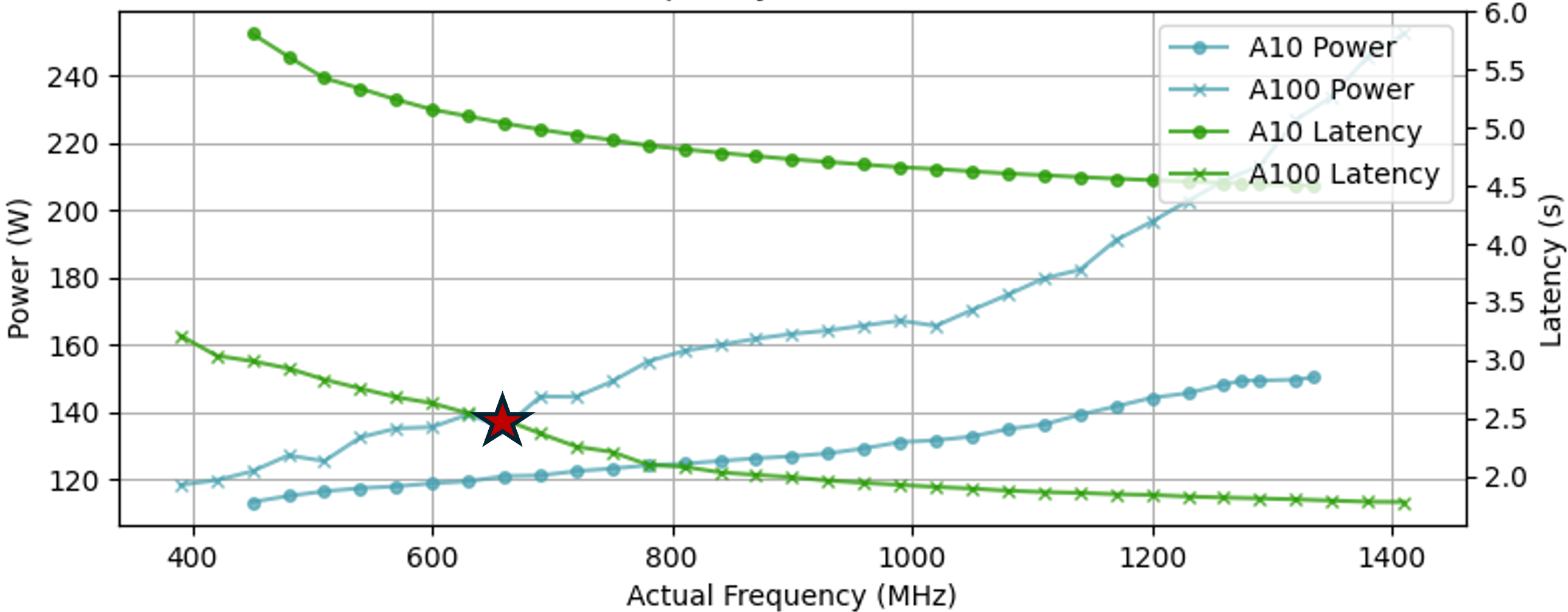}
    \vspace{-2em}
    \caption{Power-latency optimal trade-off reached at the marked point. Two GPU models are profiled as an illustration. }
    \label{fig:T1_Frequency}
    \vspace{-1em}
\end{figure}

\paragraph{Opportunity 3: Power-Latency tradeoff via frequency scaling.} Under the optimal GPU instance configurations for specialized request type and load partitioning, we can exploit the GPU frequency-throughput characteristics for each instance to further adjust GPU frequency for power minimization through DVFS in a dynamical approach~\cite{yu2025voltanallm, kakolyris2024slo}, which is adaptable to real-time LLM workload. More interestingly, such a problem draws a connection to the broader challenge of fair resource allocation among constrained workloads~\cite{low2002understanding, qiu2024power}. 

\begin{figure}[tb]
    \centering
    \includegraphics[width=1\linewidth]{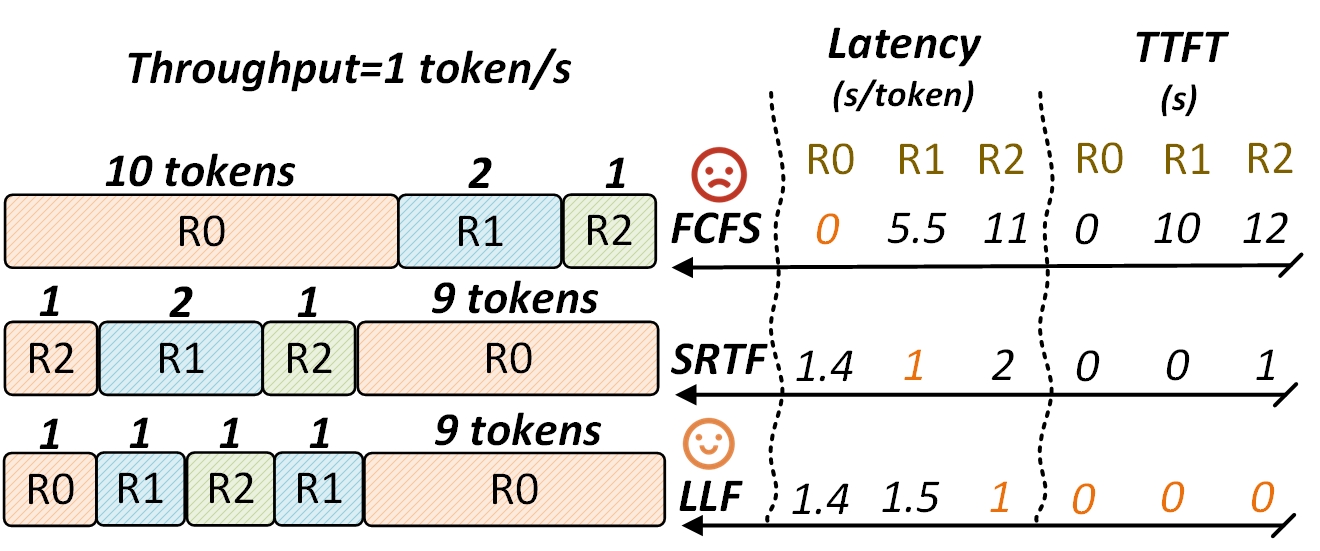}
       \caption{Toy example illustrates three requests $R_0$, $R_1$, and $R_2$ arriving sequentially at $t = 0$,$1$, and $2$, each requiring 10, 2, and 1 tokens, respectively, with a throughput of 1 token per second and scheduling policy update every second. In FCFS, $R_0$ occupies the worker entirely, causing head-of-line blocking for $R_1$ and $R_2$. In contrast, LLF dynamically prioritizes requests with smaller laxity values, allowing more urgent requests to proceed earlier. As a result, LLF achieves lower average latency, lower TTFT, and better fairness across concurrent requests compared with FCFS.}
    \label{fig:toyexample}
    \vspace{-1em}
\end{figure}

\paragraph{Opportunity 4: Fairness-Latency tradeoff via Request Laxity.} 
Achieving both LLM serving performance and fairness in a power-efficient way is challenging~\cite{sheng2024fairness, srivatsapreble, qiu2024efficient}. While LLM serving is non-deterministic, we identify  the server-side and query-side heterogeneity: i). inference server instances are varying over geographical locations and power profiles; ii). LLM workloads are highly diverse in their energy-performance profiles. By including a lightweight prediction model, we can define and quantify the \textit{laxity} (Sec. \ref{sec:laxity}) to represent each query's urgency, and integrate the LLF scheme with queries routed to each power-efficient TP instance. LLF avoids HOL blocking commonly seen in FCFS policy. Fig. \ref{fig:toyexample} illustrates an example with queue updated every 1 second, where given predicted request output length, LLF ensures smaller TTFT and improved fairness compared to Shortest-Remaining-Time-First (SRTF).
\section{FREESH Design}
In our proposed framework, we consider the setup of multiple model replicas which can be served on multiple GPU clusters in a distributed manner. We now consider all GPU instances are used to serve single AI/LLM model, while our MILP formulation and energy saving strategies can easily generalize to multiple LLM models with varying types of tasks~\cite{jiang2025demystifying}.
 \begin{figure*}[thb]
      \centering
      \includegraphics[width=0.8\linewidth]{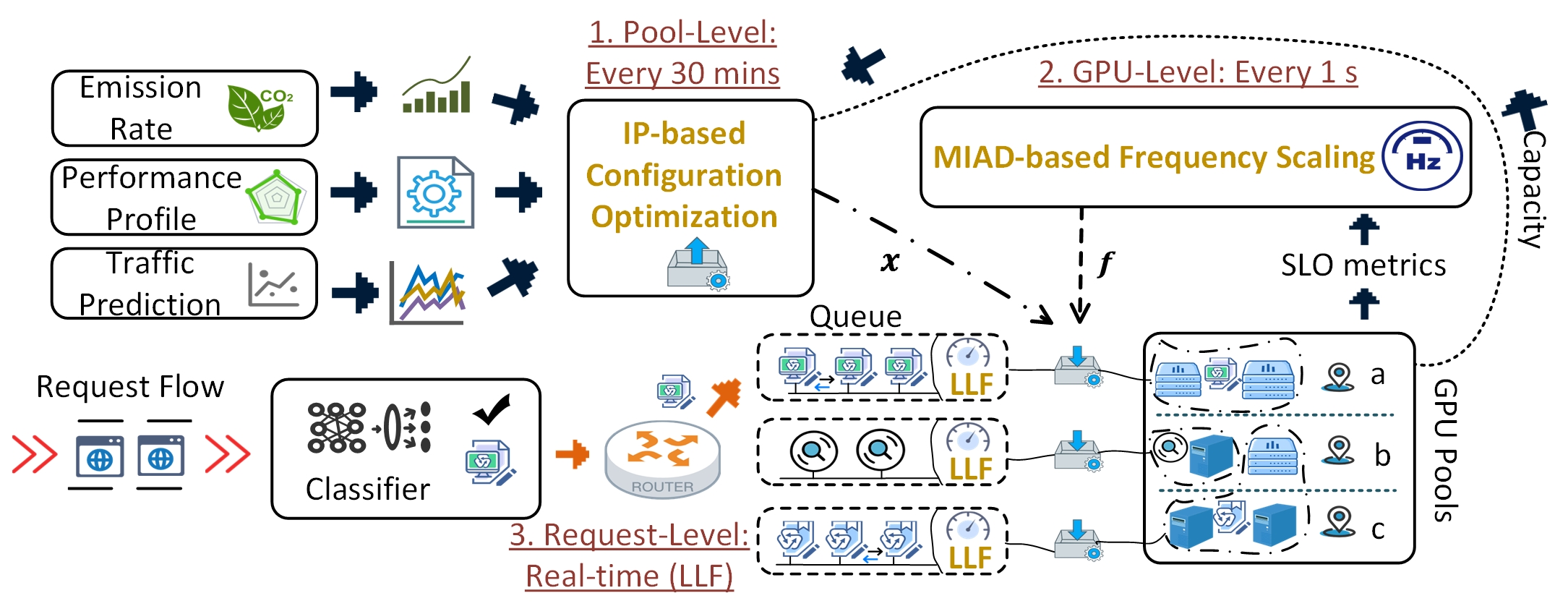}
      \vspace{-1em}
      \caption{\footnotesize Workflow of proposed FREESH for LLM serving. Here, a three-level approach is developed enabling joint optimization of emission, energy, and SLOs objectives. First, an integer programming (IP) model dynamically assigns GPUs from different locations to virtual pools every 30 minutes, with each pool tailored to a specific request category based on time-varying emission rates, performance profiles, and traffic forecasts. Incoming requests are then classified, routed to a specific queue, and scheduled for serving by a LLF algorithm that prioritizes fairness. During response generation, an MIAD algorithm dynamically adjusts the GPU's clock frequency every second to balance the trade-off between energy consumption and SLO performance. }
      \vspace{-1em}
      \label{fig:workloadflowofFREESH}
    \end{figure*}
    
\subsection{Pool-Level Cost-Aware Optimization}
The slow-timescale, pool-level decision is to identify the optimal cluster configuration to minimize operational carbon or the total electricity energy while satisfying SLO and cluster hardware constraints. Consider a group of servers with known location $n\in \mathcal{N}$ and GPU type $c\in \mathcal{C}$. $\mathcal{S}, \mathcal{R}, \mathcal{M}$ are the known sets of request partition $s$ and the corresponding request type $r$, work mode $m$, which can be specified by a LLM service operator. Together, these introduce the control \textbf{Knob 1} to \textbf{Knob 4} for LLM routing and scheduling (see Table \ref{tab:knob_summary}). The operational carbon associated with LLM serving is proportional to power consumption $\tilde{P}_{s,r,m}$ and carbon intensity or cost $\lambda_n(t)$. In this work, we propose to use locational marginal emission (LME) rate for $\lambda_n(t)$, which reflects the marginal contribution of emissions by serving GPU load at location $n$. Note $\lambda_n(t)$ can be smaller than 0 due to grid conditions~\cite{chen2024contributions}. And both historical and forecasted LME can be fetched from power grid system operators~\cite{PJM2024LME,wang2014locational}. Define $\mathcal{I} := \mathcal{N}\times\mathcal{S}\times\mathcal{R}\times\mathcal{M},\;
i=(n,s,r,m)\in\mathcal{I}$. For an index $i\in\mathcal{I}$, let $n(i),s(i),r(i),m(i)$ denote the corresponding components.
Also define index-subsets $
\mathcal{I}_r:=\{\,i\in\mathcal{I}: r(i)=r\,\},\quad
\mathcal{I}_n:=\{\,i\in\mathcal{I}: n(i)=n\,\},\quad
\mathcal{I}_s:=\{\,i\in\mathcal{I}: s(i)=s\,\}.$ We introduce compact parameter notation $ \tilde P_i := \tilde P_{s,r,m}$, $\bar L_i := \bar L_{s,r,m}$, and $G_{m,c}$ kept as-is (applied to $m(i)$ when needed). Then the optimization of $x_i$ can be formulated as an integer program (IP):
\begin{small}
\begin{subequations} \label{Upper-IP}
\begin{align}  
 \min_{x_{i}\in\mathbb{N}} \quad
& \sum_{t\in\mathcal T}\sum_{i\in\mathcal I} \lambda_{n(i)}(t)\,\tilde P_{i}\,x_{i}\,\Delta t \label{upper_obj} \\
 \text{s.t.} \quad
& \sum_{i\in\mathcal I_{r}} x_{i}\,\bar L_{i} \;\ge\; \alpha\,\tilde L_{r,t}
\quad\forall r\in\mathcal R,\; \forall t\in\mathcal T, \\
 & \sum_{i\in\mathcal I_{n}} x_{i}\,G_{m(i),c} \;\le\; \bar G_{n,c}
\quad\forall n\in\mathcal N,\; \forall c\in\mathcal C, \\
 & \sum_{i\in\mathcal I_{s}} x_{i} \;\le\; q_{s}\;\sum_{i\in\mathcal I} x_{i}
\quad\forall s\in\mathcal S, \\
 & x_{i}\,\mathrm{SLO}_{s(i),r(i),m(i)} \;\le\; \boldsymbol{I}_{\mathbb{N}^{+}}(x_{i})\overline{\mathrm{SLO}}_{r(i)}
\quad\forall i\in\mathcal I, \\
 & \sum_{s\in\mathcal S} q_{s}=1,\qquad q_{s}\in\{0,1\} , \; \forall s.
\end{align}
\end{subequations}
\end{small}
where equation (\ref{upper_obj}) aims to minimize the carbon emissions or cost for the upcoming episode. The variable $x_{n,s, r,m}$ denotes the number of workers stationed at location $n$, assigned to request type $r$ under the partition way $s$, and operating under work mode $m$. $\Tilde{\text{P}}_{s, r,m}$ corresponds to the operating power consumption for workers configured under $\textbf{Config}\{s,r,m\}$. $\text{G}_{m,c}$ indicates the number of GPUs of type $c$ required for work mode $m$; for example, a worker operating under the mode $\hat{m} =$ (Prefill-TP2-[$c$=A100]-Decode-TP4-[$c$=A6000]) would have $\text{G}_{\hat{m},A100} = 2$ assigned for prefill phase and $\text{G}_{\hat{m},A6000} = 4$ assigned for decoding phase~\cite{zhong2024distserve}, where ``TP2'' indicates 2 GPUs are used as 1 Tensor Parallelism instance. $\overline{\text{L}}_{s,r,m}$ represents the maximum number of requests the arriving rate less than service rate achieved at the highest frequency, while $\Tilde{\text{L}}_{s,r,t}$ denotes the predicted throughput for request type $r$ at time $t$. The parameter $\alpha$ controls the risk level associated with potential throughput underestimation. $\overline{\text{G}}_{n,c}$ signifies the number of available GPUs of type $c$ at location $n$, with updates occurring in each epoch. $q_s$ denotes whether or not one partition configuration is selected. Finally, $\text{SLO}_{s,r,m}$ denotes the measured average latency for $\textbf{Config}\{s,r,m\}$, and $\overline{\text{SLO}}_{r}$ represents the latency tolerance for request type $r$ (also subject to epoch-wise updates). 

We note that such a formulation \eqref{Upper-IP} is tractable with computation time recorded in Section \ref{sec:Sensitivity}. Moreover, while we profile and assume GPU working at highest frequency to set the constraint $\overline{\text{L}}_{s,r,m}$ given request type, request partition, and work mode, this helps set an upper limit, and we further reduce emissions or enery costs using techniques described in Section  \ref{sec:MIAD2}.

\begin{table}[tb]
\vspace{-1em}
  \centering
  {\fontsize{7.5}{9}\selectfont
\setlength{\tabcolsep}{0.002pt}
  \caption{Knobs in IP-based Heterogeneous GPUs Allocation.}
\label{tab:knob_summary}
\begin{tabular}{c|c|c}
\hline
\textbf{Knob}                                 & \textbf{Sign}                              & \textbf{Examples}                                                                            \\ \hline
1.Worker Location                    & $n \in \mathcal{N}$                  & $\mathcal{N}= \{n_1: \text{Tokyo}, n_2: \text{Beijing}\}$                     \\ \hline
\multirow{2}{*}{2.Request Partition} & \multirow{2}{*}{$s \in \mathcal{S}$} & $\mathcal{S}=\{s_1:\{\text{short}: (0, 500], \text{long}: [500, \infty]\},$ \\
                                     &                               & $s_2=\{\text{short}: [0, 200], \text{long}:(200, \infty]\}\}$             \\ \hline
3.Request Type                       & $r \in \mathcal{R}_{s}$              & $\mathcal{R}_{s_1} = \{r_1:\text{short}, r_2:\text{long}\}$                   \\ \hline
4.Worker Mode                        & $m \in \mathcal{M}$                  & $M =\{m_1: \text{TP1}, m_2: \text{TP2}, m_3: \text{TP4}\}$                    \\ \hline
\end{tabular}}
\vspace{-2em}
\end{table}

\subsection{GPU-Level  Energy-Efficient Control}\label{sec:MIAD2}

\textbf{Knob 5: MIAD-based Frequency Scaling.} Built upon modern GPU DVFS capabilities, in real-time query serving, to set each TP instance's running frequency, we propose an adaptive frequency scaling mechanism based on the multiplicative increase, additive decrease (MIAD) principle~\cite{mo1999analysis}.The rationale is to capture and adapt the dynamic nature of real-time service requests in a fast-timescale, while circumventing the need for cumbersome operational data collection/fitting and complex operating point search. 
At each timestep for the LLM request $i$, denote the token generation throughput it receives as $y_i$, with an associated utility function $U_i(y_i)$. The power consumption of the GPU is a function of its operating frequency, $P(f)$ (See Fig. \ref{fig:T1_Frequency} and Fig. \ref{fig:gpu-comparison} in Appendix), while the maximum throughput $r(f)$ is another function of GPU frequency. Then to decide optimal GPU frequency, we can formulate the  following utility maximization problem subject to service rate constraints and GPU frequency limits $\underline{f}, \overline{f}$:
\begin{small}
\begin{equation}
\label{equ:utility_max}
\max_{y_i\geq0, \; \underline{f}\leq f \leq \overline{f}}\; \;   \sum_{i}U_{i}(y_i)-\beta{P({f})}  \quad 
\text{s.t.} \; \sum_i y_i \leq r(f) : \gamma. 
\end{equation}
\end{small}
where $\beta$ is a cost coefficient and $\gamma$ is the dual variable associated with the constraint, which represents the marginal value of increasing the token generation rate. Let $[\cdot]_+$ and $[\cdot]_{[min,max]}$ denote the projection operator to nonnegative scalar and box interval, respectively. The following Proposition guides our frequency scaling update:

\begin{proposition}
For the utility maximization problem \eqref{equ:utility_max}, with concave, continuously differentiable $U_i(x_i)$ and convex, continuously differentiable $P(f)$, the following condition holds for the optimal GPU frequency $f^*$:
\begin{align}
P'(f^{*}) = \frac{\gamma}{\beta}r'(f^{*});
\end{align}
Implementing the following primal-dual update rule will converge to the global optimum $f^*$ with given stepsize $k_x, k_f, k_r$: 
\begin{small}
\begin{subequations}
    \label{equ:primal-dual}
\begin{align}
y_i^{(t+1)} 
&= \Big[\, y_i^{(t)} + k_x \big( U_i'(y_i^{(t)}) - \gamma^{(t)} \big) \,\Big]_+,
\quad i=1,\dots,N, \\
f^{(t+1)} 
&= \Big[\, f^{(t)} + k_f \big( -\beta P'(f^{(t)}) + \gamma^{(t)} r'(f^{(t)}) \big) \,\Big]_{[\underline{f},\,\overline{f}]}, \\
\gamma^{(t+1)} 
&= \Big[\, \gamma^{(t)} + k_\gamma \big( \sum_{i=1}^N y_i^{(t)} - r(f^{(t)}) \big) \,\Big]_+.
\end{align}
\end{subequations}
\end{small}

\end{proposition}

See Proof in Appendix \ref{sec:MIAD} and \cite{wang2006application, low2002understanding}. Essentially,  $f$ should increase when the marginal benefit of higher generation rate $\gamma r'(f)$ exceeds the marginal cost of increased power consumption $\beta{P}'(f)$. A discrete-time, distributed algorithm can be further derived as follows given predefined sequential serving of query $i$:
\begin{small}
\begin{subequations}
\label{eq:miad_alg}
\begin{align}
&\text{Multiplicative Increase (MI):~}  f(t+1) = \min(Mf(t),\overline{f}), \\
&\text{Additive Decrease (AD):~}  f(t+1) = \max(f(t)-\delta,\underline{f}).
\end{align}
\end{subequations}
\end{small}

The \textbf{Multiplicative Increase} rule is applied when the system is under high load or congestion, thus FREESH scales frequency up rapidly to meet SLO requirements. Conversely, the \textbf{Additive Decrease} rule  is applied when the system has excess compute, gradually lowering the frequency by a fixed amount ($\delta$) to conserve power. This adaptive, closed-loop approach ensures the system consistently operates close to the optimal settings, dynamically balancing LLM serving performance with energy efficiency.

\subsection{Request-Level Real-Time Fair Scheduling} \label{sec:laxity}   

\textbf{Knob 6: LLF-based Queuing}. As above levels are request- and client-agnostic, we propose LLF scheme to schedule LLM requests which promote both fairness and efficiency at the same time. FREESH determines the priority of each request by dynamically calculating its ``laxity'' ($\ell$), a metric that quantifies the urgency of a task. The laxity representation better represents request's priority compared to FCFS or SRTF.
For a request $i$ arriving at time $t_{i, \text{arrival}}$, its laxity $\ell_i(t)$ at timestamp $t$ is defined as:
\begin{small}
\begin{equation}
\label{eq:lax}
    \ell_i(t) = (t_{i, \text{arrival}} + W_i) - t - T_{i,\text{remain}}.
\end{equation}
\end{small}
In this formula, $T_{i, \text{remain}}$ represents the predicted remaining processing time needed for request $i$ (See prediction model details in \ref{sec:prediction}), while $(t_{i, \text{arrival}} + W_i)$ constitutes the request's internal deadline. The scheduling window $W_i$ is dynamically calculated based on the request's desired latency, $Lat$.
Specifically, the client-side scheduling window is set to $W_i =\alpha_ {LLF} \times Lat$ with $\alpha_{LLF}$ a tunable parameter for adjusting window size, which serves as a relatively stringent internal time limit.
$Lat$ represents the theoretical execution duration of a request assuming no queuing delay for current GPU instance, and is calculated as $Lat = (N_{\text{tokens}} \times T_{\text{TBT}}) + T_{\text{TTFT}}$, 
where $N_{\text{tokens}}$ is the total number of predicted output tokens, $T_{\text{TBT}}$ is the Time-Between-Tokens, and $T_{\text{TTFT}}$ is the Time-To-First-Token. $T_{\text{TBT}}$ and $T_{\text{TTFT}}$ are profiled based on current GPU configurations. A request type forecasting model is applied to determine expected output length $N_{\text{tokens}}$ based on input tokens~\cite{qiu2024efficient, fu2024efficient}.
It is essential to distinguish the strict internal scheduling window from the more lenient server-side Service Level Objective (SLO), which is set to $5 \times Lat$ in this paper\cite{stojkovic2025dynamollm}.

At each timestep $t$, for every arrived and unfinished request we update $\ell_i(t)$ by \eqref{eq:lax} with minor computation costs. Then, the LLF algorithm dispatches the requests with the smallest laxity values to the available workers. Our implementation is \emph{preemptive}: if a newly arrived or updated request gets a strictly smaller $\ell$ than a running one, we may preempt the latter and reassign service to the former. 

\section{Experiments}

In this Section, our experiments are designed to answer following research questions: \textbf{1). \emph{Comparative performance.}} Can FREESH reduce carbon emissions with competitive energy efficiency and SLO attainment? \textbf{2). \emph{Controller contribution.}} What is the individual contribution of each control knob in FREESH to improving the overall performance of LLM serving? \textbf{3). \emph{Robustness analysis.}} How does FREESH perform under various serving conditions?
    
\subsection{Evaluation Setup}

\paragraph{Implementation.}
We instantiate the FREESH framework using a combination of established techniques and custom-built tools (detailed in the Appendix). It utilizes \emph{vLLM} \cite{kwon2023efficient} as the inference engine for LLM serving and leverages the \emph{PuLP} library \cite{pulp} to model and apply Gurobi \cite{gurobi}  to solve IP optimization problems. Traffic prediction is handled by an \emph{LSTM} model, while the \emph{nvidia-smi} tool is used to capture power and energy characteristics and for implementing the frequency scaling. For request classification, we trained a \emph{BERT} model. The router, the LLF queue, and the communication interfaces between the different modules are all implemented in Python. This modular design allows FREESH to seamlessly integrate state-of-the-art or user-specified techniques.

\paragraph{Resources.} The experiments leverage the Llama3-70b model \cite{llama3modelcard} deployed on worker modes TP2, TP4, and TP8 on A100 or H100 GPUs with several datasets, including: (1) spatiotemporal carbon intensity profiles from PJM \cite{PJMDataMiner} (2) prompts and responses for request generation and classification from LMSys-Chat-1M \cite{zheng2023lmsyschat1m} and (3) conversation traces for workload prediction and online service traffic \cite{stojkovic2025dynamollm}. We design two distinct partitioning sets of varying partition values to classify the requests into four types, respectively, denoted by $\mathcal{R}_s=\{r_1: \text{SS}, r_2:\text{SL}, r_3:\text{LS}, r_4:\text{LL}\}$~\footnote{Our method is flexible with the set of partitions, and is scalable to more request types for finer-grained classification and partition.}, where SS indicates a request with Short input and Short output (See Table \ref{Patition} in Appendix).

\paragraph{Metrics.} FREESH aims to achieve a trade-off between energy, environmental effects, and SLO attainment. To quantify these, our environmental metrics include energy consumption and carbon emissions. For evaluating SLOs, we employ Time-to-First-Token (TTFT), Time-Between-Token (TBT), Violation Rate, and Jain Fairness \cite{jain1984quantitative}.  (See Table \ref{tab:slo_thresholds} in Appendix for metric thresholds)

\paragraph{Testbeds.} To establish our testbeds, we assume three distinct locations, each characterized by a different carbon intensity. The geographically distributed servers are configured as follows: 4 A100s and 4 H100s at location 1, 12 A100s and 2 H100s at location 2, and 8 A100s at location 3. Then, we selected two consecutive episodes (Timeslot 24 and 25) from the whole 48 episodes of 1-day pool-level optimization shown in Fig. \ref{T1_mode_withFREESH} to demonstrate results. These episodes are denoted as $\texttt{E1}$ and $\texttt{E2}$, respectively, with each comprising 6 time steps, representing 30 minutes. Furthermore, the workload level and spatiotemporal carbon intensities differ between $\texttt{E1}$ and $\texttt{E2}$, while all other parameters remain identical, providing a just comparative basis.


\paragraph{Baselines.} Initially, we conduct 1-day simulations and compare our FREESH with the \texttt{MinEnergy} minimizing energy consumption at pool-level. For fast-timescale operation performance, we compared with \texttt{Baseline} which is without the request classifier nor any of the proposed control knobs. To demonstrate the impact of individual control knob:  request partition, work mode shifting, frequency scaling, and LLF scheduling, we conduct shorter, 1-hour experiments as ablation studies involving \texttt{w/o MultiParti}, \texttt{w/o TPchange}, \texttt{w/o MIAD} and \texttt{w/o LLF} where 'w/o' means the implementation of FREESH without the corresponding knob. We also compare the proposed LLF with the classic queuing strategies of Early-Deadline-First (\texttt{EDF}), Shortest-Remaining-Time-First (\texttt{SRTF}) and First-Come-First-Serving (\texttt{FCFS}).

\subsection{Performance Analysis}
\begin{figure*}[ht] 
      \centering
      \vspace{-0.5em}
\includegraphics[width=0.95\linewidth]{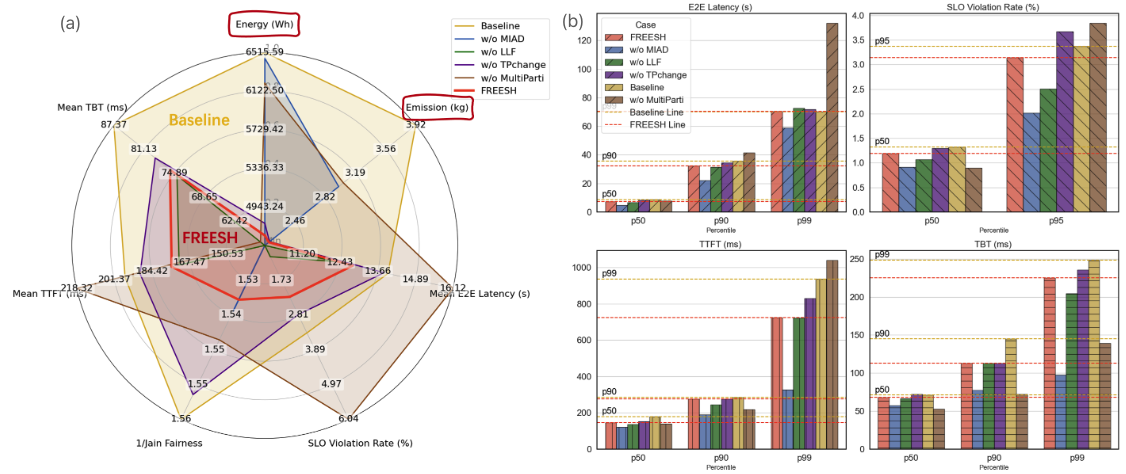}
      \vspace{-1em}
      \caption{Performance comparison on a. (Left) 7 metrics with control knobs variations; and b. (Right) under varying percentiles. } \label{baseline}
      \vspace{-1em}
    \end{figure*}
    
\begin{table*}[tbh]
\centering
\caption{Sensitivity of LLF to workload level compared to FCFS. Shaded entries denote higher performances.}\label{tab: LLF_sensi}{\fontsize{8}{10}\selectfont
\setlength{\tabcolsep}{4pt} 
\begin{tabular}{c|ccccccc}
\hline
Metrics    & Energy (Wh)                                & Emission (kg)                           & SLO Vio. Rate                           & Mean Latency (s)                          & Mean TTFT (ms)                           & Mean TBT (ms)                           & Jain Fairness                          \\ \hline
LLF        & 4651.81                                    & 2.14                                    & 0.02                                    & 12.85                                     & 175.53                                   & 75.79                                   & 0.65                                   \\
FCFS       & \cellcolor[HTML]{ECF4FF}4550.15 (-2.18\%)  & \cellcolor[HTML]{ECF4FF}2.09 (-2.37\%)  & \cellcolor[HTML]{ECF4FF}0.01 (-60.05\%) & \cellcolor[HTML]{ECF4FF}12.28 (-4.41\%)   & \cellcolor[HTML]{ECF4FF}172.17 (-1.91\%) & \cellcolor[HTML]{ECF4FF}74.33 (-1.93\%) & \cellcolor[HTML]{ECF4FF}0.66 (+0.70\%) \\ \hline
LLF-2xQPS  & \cellcolor[HTML]{DAE8FC}4653.26 (-9.35\%) & \cellcolor[HTML]{DAE8FC}2.15 (-9.40\%) & \cellcolor[HTML]{DAE8FC}0.35(-70.47\%) & \cellcolor[HTML]{DAE8FC}109.55(-62.77\%) & 335.28                                   & 77.46                                   & \cellcolor[HTML]{DAE8FC}0.65(+4.84\%)  \\
FCFS-2xQPS & 5133.22                                    & 2.38                                    & 0.94                                    & 370.98                                    & \cellcolor[HTML]{ECF4FF}310.80 (-7.30\%) & \cellcolor[HTML]{ECF4FF}76.07 (-1.79\%) & 0.62                                   \\ \hline
\end{tabular}}
\end{table*}

\begin{figure} [thb]
      \centering
\includegraphics[width=1\linewidth]{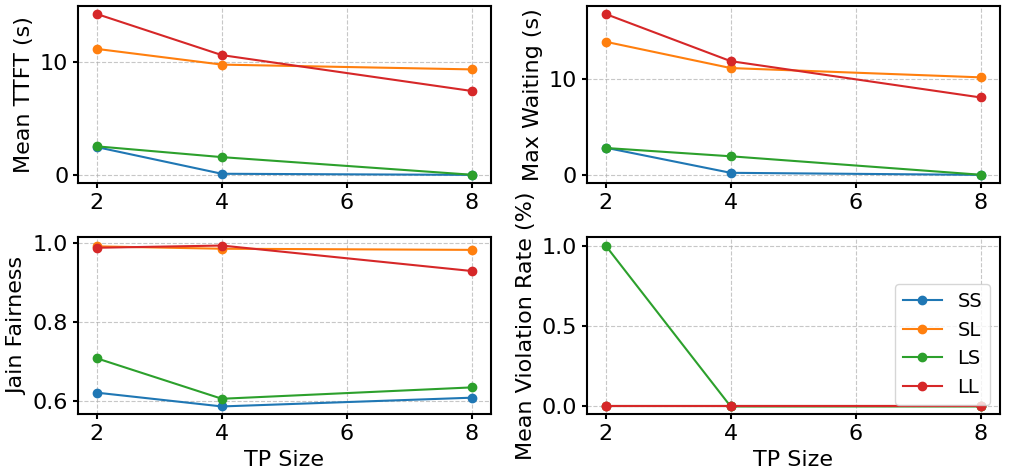} 
      \caption{Sensitivity of LLF performance to TP mode changes.} \label{llf_gpu_main}
    \end{figure}

 \begin{figure*}[htb] 
      \centering
      \includegraphics[width=1\linewidth]{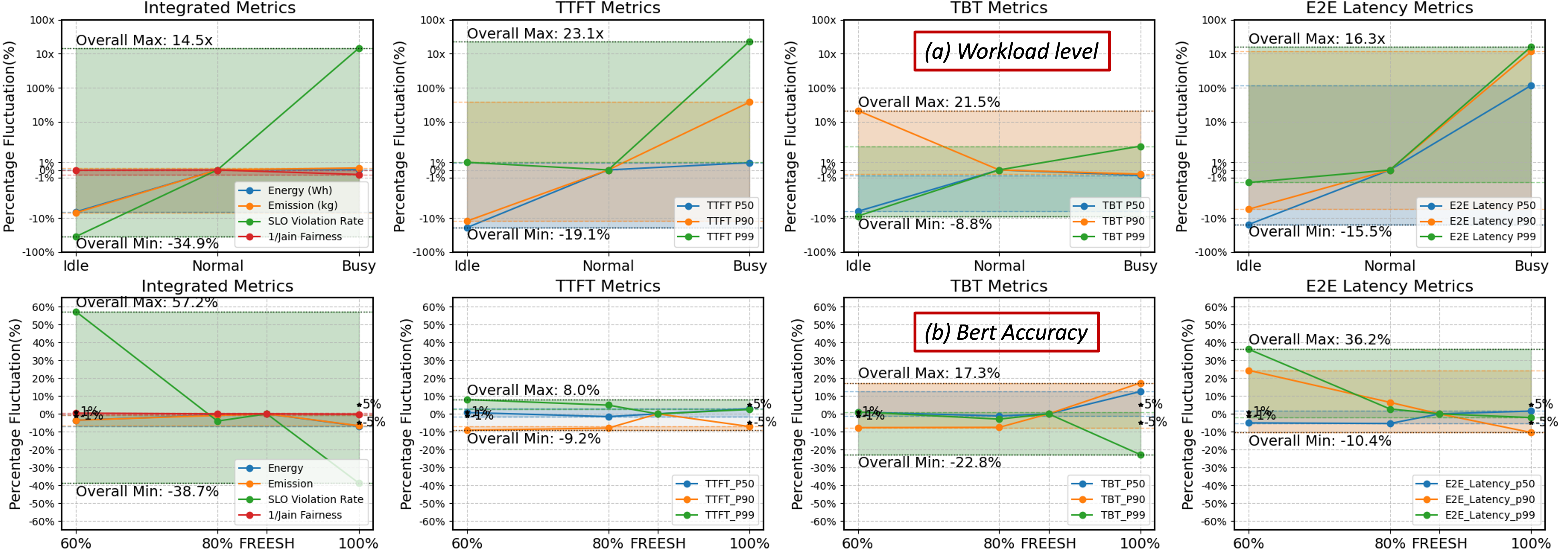}
      \caption{The sensitivity of Cost and Performance metrics to workload level (Top) and Bert accuracy (Bottom).}\label{sensitivity}
    \end{figure*}
    
\paragraph{Energy and Emission.} Compared to \texttt{Baseline}, our FREESH achieves substantial reductions by leveraging request-type information and efficient scheduling: an 28.60\% (1863.78 Wh) energy reduction and a 45.45\% (1.78 kg) emission reduction under real emission and workload scenarios. This is due to the looh-ahead GPU cluster provisioning, adaptive DVFS, and energy-efficient request partitioning. Fig. \ref{baseline}(a) indicates that all optimization knobs except LLF contributed positively to these results. This suggests that LLF's benefit is limited under the traffic conditions of episodes \texttt{E1} and \texttt{E2}, while LLF is more beneficial for LLM serving fairness.


\paragraph{SLO and Fairness.} As detailed in Fig. \ref{baseline} (a) and Fig. \ref{baseline} (b), FREESH delivers notable latency improvements across multiple metrics compared to \texttt{Baseline}. The E2E latency sees reductions in the mean (8.20\%), P50 (16.41\%), and P90 (9.07\%), though the P99 latency exhibited a marginal 0.34\% increase. TTFT shows stronger gains, particularly at P50 (18.86\%) and P99 (22.77\%). Similarly, TBT is reduced significantly, with the P50 and P99 demonstrating reductions of 21.98\% and 13.25\%, respectively. Given the absence of congestion, all configurations maintained an acceptable and comparable level of fairness and SLO violation rate, with FREESH ultimately performs best.

\paragraph{Cost and Performance Trade-off.}
As shown in Fig. \ref{baseline} (a), the individual optimization knobs within FREESH present distinct performance effects (gain and debuff). The strategic coordination of these knobs is the primary mechanism enabling the integrated FREESH system to outperform the Baseline across all metrics. Furthermore, while the LLF knob provides a congestion mitigation capability intended for high-load scenarios, its strategic inclusion ensures system robustness (Verified in Section \ref{Workload_sec}). This coordinated design demonstrates FREESH's effectiveness in achieving a superior balance among the competing objectives of cost, SLO, and fairness in LLM service.

\subsection{Ablation Studies}

\paragraph{Request partition.} When fixed to a non-optimal request partitioning, the system's resource allocation, optimized by model $\ref{Upper-IP}$, can be altered to include higher-capacity configurations (e.g., from TP1-A100 to TP4-H100/TP2-A100). As shown in Fig. \ref{baseline}, this change to fixed partitioning causes visible, more idle resource, leading to higher energy consumption (+33.32\%) and increased emissions (+42.26\%) due to over-provisioning. While the excess hardware capacity can improve mean, P50, P90 and P99 TBT (-24.70\% to -38.14\%), the non-optimal request partition causes an imbalanced combination of short and long requests, which manifests as a sharp increase in P90 (+28.60\%) and P99 (+87.67\%) E2E latency, and P99 (+43.86\%) TTFT, indicating the presence of HOL congestion for certain requests. This observation demonstrates that a poor partition can perform worse than \texttt{Baseline}, overriding the benefits of sophisticated hardware configuration optimization and confirms that optimal request partitioning is essential for achieving load balance and improving SLO attainment.

\vspace{-10pt}\paragraph{TP mode changes.}
When the TP mode remains fixed to be the optimal for episode $\texttt{E1}$, when the workload is transitioning to $\texttt{E2}$, a degradation is observed across both SLO and efficiency metrics. Energy consumption increases by $2.66\%$ and carbon emission by $0.40\%$. E2E latency metrics show consistent degradation. TTFT increases across key percentiles: $\text{mean}$ ($+8.11\%$), $\text{P50}$ ($+4.21\%$), and $\text{P99}$ ($+14.70\%$), with a minimal decrease at $\text{P90}$ ($-0.44\%$). Similarly, TBT rises for most of percentiles: $\text{mean}$ ($+4.02\%$), $\text{P50}$ ($+6.11\%$), and $\text{P99}$ ($+4.71\%$), with a negligible decrease at $\text{P90}$ ($-0.13\%$). Despite these increases, the mean, $\text{P50}$, and $\text{P95}$ SLO violation rates remain low, ranging from $1.4\%$ to $3.7\%$. This suggests that the workload in $\texttt{E2}$, even under the suboptimal TP mode, operates below the critical resource saturation point.

Given that other knobs inherently manage global resource allocation, their interface with multi-GPU setups is straightforward. In contrast, LLF is a purely per-request policy, which poses a potential challenge for scaling its effectiveness. To validate its multi-GPU performance, we simulate its behavior with TP mode changes. Fig. \ref{llf_gpu_main} shows that among general inference speedups achieved by scaling resources, LLF consistently maintains the highest Jain Fairness for long-output sequences. This confirms the LLF's reliability and sustained effectiveness across diverse resource pools.
    
\vspace{-10pt}\paragraph{MIAD-based frequency scaling.} Without the MIAD mechanism, the scheduling in FREESH defaults to aggressively pursuing latency, resulting in performance metrics reaching their best-observed levels at a significant cost to efficiency. The absence of MIAD scaling is evidenced by substantial improvements across all latency and SLO metrics: TTFT shows reductions up to $-54.88\%$ ($\text{P99}$), TBT is reduced by up to $-56.65\%$ ($\text{P99}$), and overall E2E latency is significantly reduced (e.g., $\text{P50}$: $-36.73\%$). This aggressive scheduling directly translates to a minimal SLO violation rate ( $-71.16\%$) and a slight change in Jain Fairness ($-0.15\%$). However, this peak performance is achieved because the system is forced into a worst-case provisioning strategy, maintaining the highest frequency to maintain the SLO attainment. Consequently, efficiency metrics suffer considerably: energy consumption increases by $38.81\%$ and carbon emission by $39.61\%$. This stark trade-off highlights that the core function of MIAD-based frequency scaling is indispensable for avoiding unnecessary idle power consumption and achieving cost-effective LLM service.


\paragraph{LLF-based scheduling.} As shown in Fig. \ref{baseline} and Table \ref{tab: LLF_sensi}, the removal of LLF scheduling improves across several metrics when operating under a normal workload level without significant congestion. This counter-intuitive degradation is attributed to the inherent design of LLF: operating under a normal workload, the scheduler’s aggressive focus on realizing the preemptive mechanism and meeting a strict client-side serving time window introduces unnecessary overhead. Since preemptive actions are easily triggered even without actual congestion, the extra time consumed by request scheduling logic hurts SLO attainment and increases energy and emission, demonstrating that LLF's benefit is more significant under high-contention scenarios (Sec \ref{Workload_sec}).
\begin{figure*}[ht]
      \centering
      \includegraphics[width=0.95\linewidth]{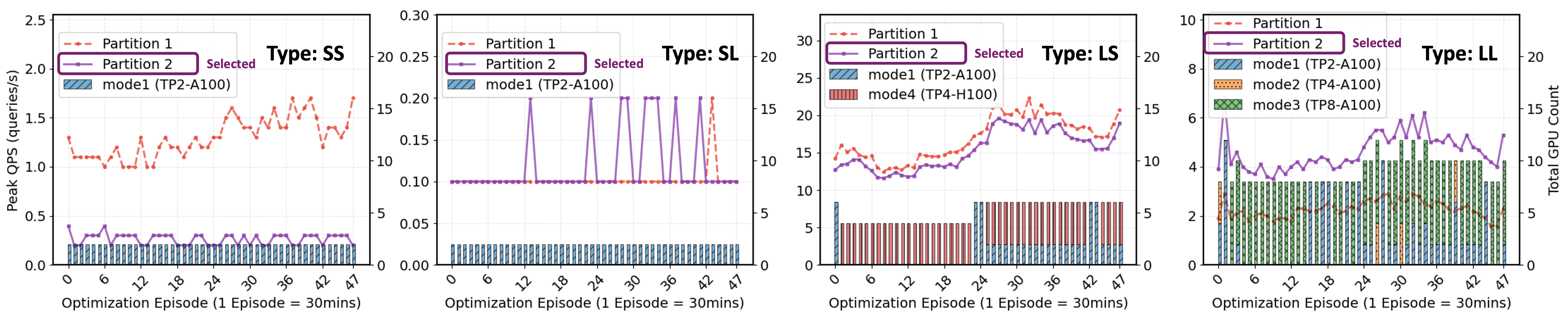}
      \caption{TP mode variation across heterogeneous GPUs across 1-day traffic for 4 request types.} \label{T1_mode_withFREESH}
    \end{figure*}
    
\begin{figure}[ht]
      \centering
      \includegraphics[width=0.99\linewidth]{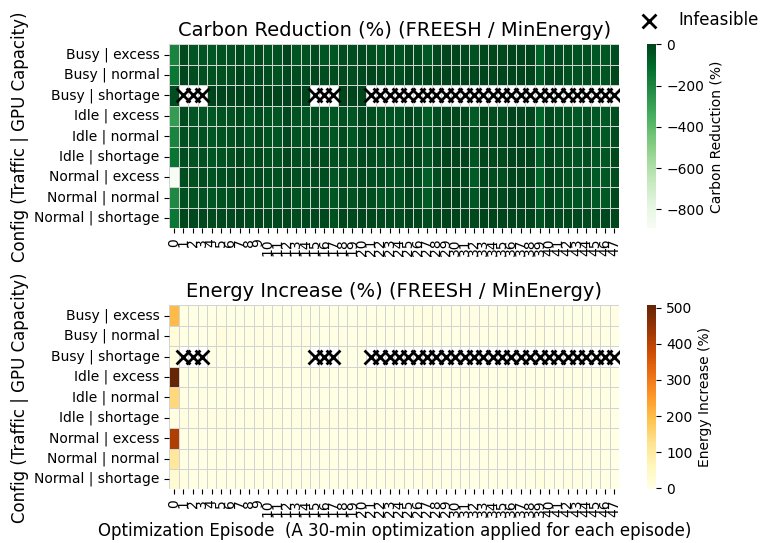} 
      \caption{Sensitivity to pool-level system parameters.} \label{sensi_gpu}
    \end{figure}
    
\subsection{Sensitivity Analysis} \label{sec:Sensitivity}
\paragraph{Workload Level. } \label{Workload_sec}
To verify the effectiveness of the LLF mechanism and demonstrate its response to potential real-time workload fluctuations, we configured a testbed at 2xQPS (Busy) while keeping other configurations unchanged. As shown in Table $\ref{tab: LLF_sensi}$, LLF-Busy achieves performs better than FCFS-Busy across most critical metrics, validating LLF's ability to manage high contention. LLF's advanced request prioritization logic results in a significant mitigation of resource contention and queuing delays: mean E2E latency is lowered by $70.47\%$ compared to FCFS-Busy along with the SLO violation rate is reduced by $62.77\%$, showcasing LLF's robust stability under high load. Further, LLF-busy achieves an energy consumption reduction of $9.40\%$ and a carbon emission reduction of $9.35\%$ compared to FCFS-Busy. The consistent improvements across efficiency, SLO, and Jain Fairness (which improves by $4.84\%$) confirm that LLF's preemptive scheduling is essential for achieving reliable and cost-effective LLM service under high-contention traffic.

To demonstrate the sensitivity of FREESH with respect to workload level, we further configured a testbed with 0.5xQPS (Idle) to test the performance of FREESH encountering the more idle workload. As shown in Fig. \ref{sensitivity} (Top), the system effectively adapts its energy footprint based on demand: energy consumption and carbon emission decrease by about $7\%$ in the Idle scenario. In contrast, under the Busy scenario, these metrics only increase by about $1\%$. This effective scaling is attributed to the MIAD mechanism, which maximizes efficiency across diverse workload levels. While Jain Fairness remains stable across all workload levels, SLO metrics deteriorate more in the Busy scenario. Although the LLF mechanism can mitigate request bursts by prioritizing urgent tasks, the benefits are likely offset by the internal FCFS scheduler within the \emph{vLLM}. 


    
\paragraph{Classifier Accuracy.} 

The incoming workload pattern is determined by our BERT request predictor and affecting FREESH decisions, making it necessary to check the sensitivity to prediction accuracy, which is illustrated in Fig. \ref{sensitivity} (Bottom). Both energy consumption and energy emission show relative stability, fluctuating only within a narrow $\pm 5\%$ band across all accuracy levels. This confirms that FREESH's adaptive component maintains resource efficiency even with moderate prediction errors. Jain Fairness also demonstrates high stability, remaining within the $\pm 5\%$ band. However, SLO metrics exhibit notable volatility. TBT shows a substantial maximum fluctuation, ranging from $-22.8\%$ to $+17.3\%$, and E2E latency is constrained between $-10.4\%$ and $+36.2\%$. This volatility is mirrored by the SLO violation rate, which exhibits the largest fluctuation, spanning from $-38.7\%$ to a maximum of $+57.2\%$. The strong performance under $100\%$ accuracy case confirms that precise request prediction is critical for reliable SLO.

\paragraph{System Parameter. } All scheduling decisions within our framework are intrinsically linked to the optimal resource configuration derived from the pool-level IP solution \eqref{Upper-IP}. While this solution depends on various system parameters, we zoom in those with most frequent fluctuation: request traffic prediction, GPU capacity, and carbon intensity. As shown in Fig. \ref{T1_mode_withFREESH}, the selection of Partition 2 causes the GPU allocation strategy closely tracking the workload's traffic fluctuations. This tight coupling confirms the high sensitivity and adaptive capacity of the GPU resource allocation within FREESH to the request traffic prediction. Fig. \ref{sensi_gpu} demonstrates that both request traffic and GPU capacity are critical factors influencing the efficiency and feasibility of FREESH. The system becomes infeasible when faced with a GPU capacity shortage combined with busy traffic. Feasibility is restored by either enlarging the GPU capacity or decreasing the traffic load, allowing FREESH to achieve significant carbon reduction compared to \texttt{MinEnergy}. In most scenarios, FREESH achieves substantial carbon reduction with only a marginal energy increase compared to \texttt{MinEnergy}. However, a notable divergence occurs at timeslot 1: the presence of negative carbon intensity, as shown in Figure \ref{fig:Spatiotemporal}, fundamentally alters the optimization landscape, causing the scheduling decisions made by \texttt{MinEnergy} and FREESH to differ significantly and resulting in a large variance in both carbon emission and energy consumption.
    

 \begin{figure} [ht]
      \centering
      \includegraphics[width=0.89\linewidth]{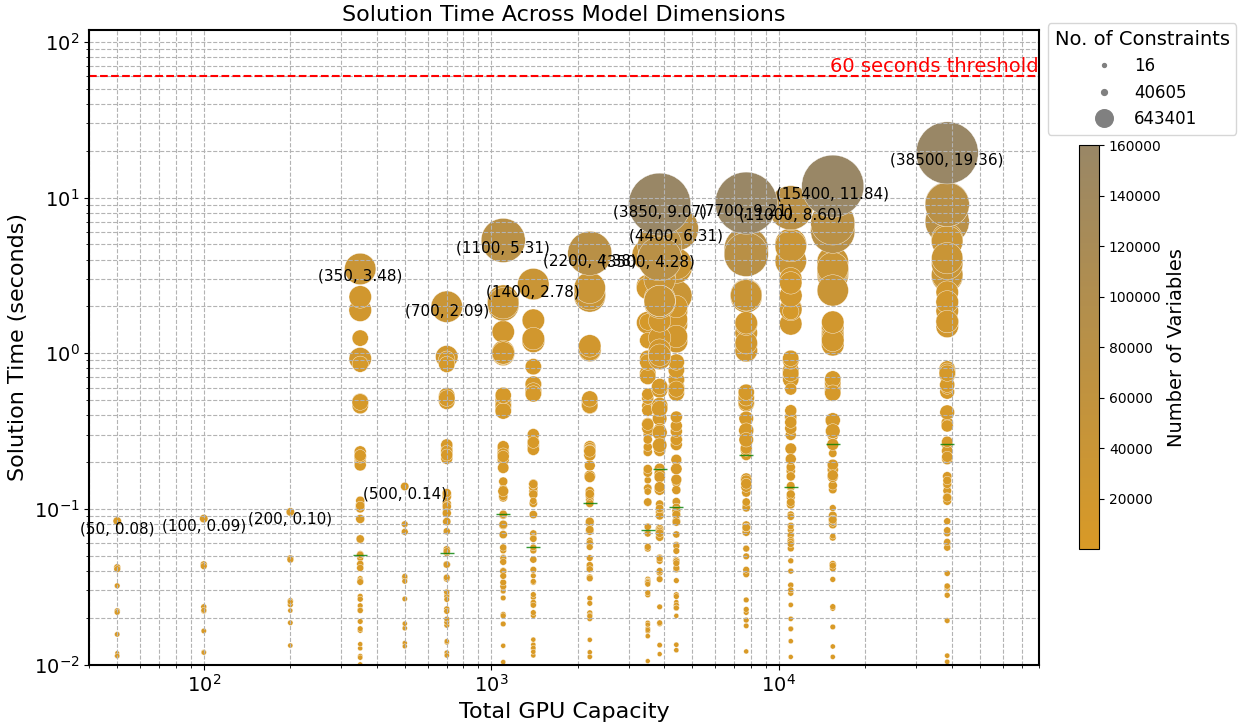} 
      \caption{Solution time across pool-level IP model scale.} \label{solu_tra}
    \end{figure}

\paragraph{System Scale. } The solution complexity of the optimization model \ref{Upper-IP} is contingent upon the system scale, while MIAD and LLF adds negligible computations. We evaluated various systems by combining different numbers of pool-level control knobs, where the solution space dimension is defined by $|\mathcal{N}|\times|\mathcal{S}|\times|\mathcal{R}|\times|\mathcal{M}|$, varying the number of knobs within the set $\{1, 5, 10, 20\}$. Figure \ref{solu_tra} indicates that the solution time is not primarily determined by the Total GPU Capacity. Instead, it mainly depends on the IP model's complexity, specifically the combination of control knobs. This decoupling suggests a crucial practical implication: it is feasible to use a relatively small IP model to configure considerable GPUs. Furthermore, the model demonstrates high efficiency, with all observed solution times remaining less than 60 seconds, which is practical for slow-timescale optimization. Even for the largest test case, where the total GPU Capacity reaches 38,500, the number of constraints is approximately 643,401, and the number of variables is around 160,000, the solution time remains only $19.36$ seconds. This rapid process confirms the practicality of solving the IP model online to realize configuration optimization.

\section{Related Work}
\label{gen_inst}

\textbf{LLM Serving} Various design objectives of LLM serving have been discussed in the literature. For instance, \cite{li2023alpaserve} considers the operator placement strategies for latency reduction. \cite{ye2025flashinfer} develops FlashInfer to tackle KV-cache storage heterogeneity using block-sparse format and composable formats to optimize memory access and reduce redundancy. It also proposes the load-balanced scheduling algorithm. Vidur models the performance of LLM operators via a combination of experimental profiling and predictive modeling, and evaluates the inference performance for different workloads by estimating SLO metrics~\cite{agrawal2024vidur}. An MILP is formulated in \cite{mei2024helix} to select a pipeline of GPU instances to place all layers of the LLM, while ours focus on routing and scheduling optimization.

In the pre-LLM serving schemes, such as Clipper \cite{crankshaw2017clipper} and Rafiki \cite{wang2018rafiki} apply the practical Additive-Increase-Multiplicative-Decrease (AIMD) algorithm to adjust the inference batch size.
While $\mu$-Serve demonstrates the benefits of GPU frequency scaling in power saving for model serving~\cite{qiu2024power}, which also inspires us to extend multiplicative-increase additive-decrease strategy on cluster-level frequency with theoretical justifications. More interestingly, we show the promises of not only reducing power but also total energy consumption or emissions associated with LLM serving. 

Research in LLM scheduling also have highlighted the potential of improving SLO attainment and fairness. FastServe adopts a multi-level feedback queue to improve the average job completion time, while the non-deterministic model executions still introduce service variability~\cite{wu2023fast}. Various strategies such as learning to rank the shortest job~\cite{fu2024efficient},  speculative SRTF scheduler with a light proxy output length prediction model~\cite{qiu2024efficient}, virtual token counter for fairness scheduling~\cite{sheng2024fairness, cao2025locality} have been proposed to optimize the queued requests. Observing the varying computation needs of prefill and decoding phases, research also proposes caching and serving schemes tailored for each phase~\cite{srivatsapreble, patel2024splitwise, agrawal2024taming}. While past research do not exploit temporal and spatial flexibility jointly~\cite{lindberg2022using, stojkovic2025dynamollm}, or rely on either optimization-based approaches, which are not tractable in real-time operation~\cite{breukelman2024carbon}, or suboptimal heuristics as they do not take into account future LLM traffic or emission signals~\cite{duan2024muxserve, jiang2025demystifying}.

\textbf{LLM's Energy and Environmental Impacts} 
Classical works centered around improving power and energy efficiency of data centers through power capping~\cite{li2020thunderbolt}, scheduling~\cite{berral2010towards}, and load balancing~\cite{liu2011greening}. With the disruptive increase on LLM inference load and intensive power density, unseen challenges arise over the carbon and energy profiles for LLM queries and AI data centers~\cite{maji2025data, li2025ecoserve}. Recent studies also identified the power and energy availability may be the future bottleneck of LLM deployments~\cite{li2024unseen, stojkovic2025tapas, MIT2025AI}. Solutions have been proposed in terms of co-planning for future data centers and power infrastructures to reduce associated carbon footprints~\cite{lin2023adapting, ferris2025optimizing}. Recent works show the promise of carbon reductions by geographically shifting data center loads to low-emission-rate regions~\cite{lindberg2020environmental, chien2022beyond}. In Clover, a carbon-friendly ML inference service runtime system is proposed which balances performance, accuracy, and carbon emissions through mixed-quality models and GPU resource partitioning, where simulated annealing is applied evolutionarily in graph-based search space \cite{li2023clover}. 

Previous research look into the cost, power and energy efficiency of LLM serving systems~\cite{li2025ecoserve, jiang2025demystifying, sukprasert2024limitations}. DynamoLLM pioneers in reconfiguring inference clusters to optimize for energy and cost of LLM serving under the service’s performance SLOs~\cite{stojkovic2025dynamollm}. DynamoLLM along with a few research identify the energy-saving DVFS and TP configuration mainly based on the discretized search space of energy-GPU configuration table~\cite{kakolyris2024slo, kakolyris2025throttll}, which can lead to suboptimal solutions. Moreover, the dynamic and autoregressive token generation process makes it challenging to precisely pre-allocate computational resources for incoming inference requests, while DynamoLLM does not consider cross-instance coordination or emission impacts. SpotHedge proposes to spread spot replicas across clouds based on costs~\cite{mao2025skyserve}, while our work is novel in considering cross-region emission/energy-guided serving optimization compatible with fair and efficient LLM scheduling.


\section{Conclusion and Future Work}
This paper identifies the joint computation and energy challenges of serving LLM workloads across geographically distributed instances. We present FREESH, a hierarchical and coordinated routing and scheduling scheme by fully leveraging the heterogeneity of locational carbon emission rate, LLM workloads and GPU instances' energy-performance characteristics. FREESH provides a novel spatiotemporal coordination framework for energy-efficient cross-region LLM serving, and it demonstrates very competitive results through tractable GPU instance optimization, adaptive frequency scaling, and LLF scheduling policy. In the future, we will consider more diverse LLM serving tasks and fine-grained partition and GPU configuration for achieving improved energy and emission savings.



\nocite{langley00}

\bibliography{bib}

\begin{thebibliography}{67}
\providecommand{\natexlab}[1]{#1}
\providecommand{\url}[1]{\texttt{#1}}
\expandafter\ifx\csname urlstyle\endcsname\relax
  \providecommand{\doi}[1]{doi: #1}\else
  \providecommand{\doi}{doi: \begingroup \urlstyle{rm}\Url}\fi

\bibitem[Agrawal et~al.(2024{\natexlab{a}})Agrawal, Kedia, Mohan, Panwar, Kwatra, Gulavani, Ramjee, and Tumanov]{agrawal2024vidur}
Agrawal, A., Kedia, N., Mohan, J., Panwar, A., Kwatra, N., Gulavani, B.~S., Ramjee, R., and Tumanov, A.
\newblock Vidur: A large-scale simulation framework for llm inference.
\newblock \emph{Proceedings of Machine Learning and Systems}, 6:\penalty0 351--366, 2024{\natexlab{a}}.

\bibitem[Agrawal et~al.(2024{\natexlab{b}})Agrawal, Kedia, Panwar, Mohan, Kwatra, Gulavani, Tumanov, and Ramjee]{agrawal2024taming}
Agrawal, A., Kedia, N., Panwar, A., Mohan, J., Kwatra, N., Gulavani, B., Tumanov, A., and Ramjee, R.
\newblock Taming $\{$Throughput-Latency$\}$ tradeoff in $\{$LLM$\}$ inference with $\{$Sarathi-Serve$\}$.
\newblock In \emph{18th USENIX Symposium on Operating Systems Design and Implementation (OSDI 24)}, pp.\  117--134, 2024{\natexlab{b}}.

\bibitem[AI@Meta(2024)]{llama3modelcard}
AI@Meta.
\newblock Llama 3 model card.
\newblock 2024.
\newblock URL \url{https://github.com/meta-llama/llama3/blob/main/MODEL_CARD.md}.

\bibitem[Alizadeh \& Jadid(2015)Alizadeh and Jadid]{alizadeh2015dynamic}
Alizadeh, B. and Jadid, S.
\newblock A dynamic model for coordination of generation and transmission expansion planning in power systems.
\newblock \emph{International Journal of Electrical Power \& Energy Systems}, 65:\penalty0 408--418, 2015.

\bibitem[Berral et~al.(2010)Berral, Goiri, Nou, Juli{\`a}, Guitart, Gavald{\`a}, and Torres]{berral2010towards}
Berral, J.~L., Goiri, {\'I}., Nou, R., Juli{\`a}, F., Guitart, J., Gavald{\`a}, R., and Torres, J.
\newblock Towards energy-aware scheduling in data centers using machine learning.
\newblock In \emph{Proceedings of the 1st International Conference on energy-Efficient Computing and Networking}, pp.\  215--224, 2010.

\bibitem[Boyd \& Vandenberghe(2004)Boyd and Vandenberghe]{boyd2004convex}
Boyd, S.~P. and Vandenberghe, L.
\newblock \emph{Convex optimization}.
\newblock Cambridge university press, 2004.

\bibitem[Breukelman et~al.(2024)Breukelman, Hall, Belgioioso, and D{\"o}rfler]{breukelman2024carbon}
Breukelman, E., Hall, S., Belgioioso, G., and D{\"o}rfler, F.
\newblock Carbon-aware computing in a network of data centers: A hierarchical game-theoretic approach.
\newblock In \emph{2024 European Control Conference (ECC)}, pp.\  798--803. IEEE, 2024.

\bibitem[Cao et~al.(2025)Cao, Wang, Mao, Hsu, Yin, Xia, Li, Liu, Zhang, Zhou, et~al.]{cao2025locality}
Cao, S., Wang, Y., Mao, Z., Hsu, P.-L., Yin, L., Xia, T., Li, D., Liu, S., Zhang, Y., Zhou, Y., et~al.
\newblock Locality-aware fair scheduling in llm serving.
\newblock \emph{arXiv preprint arXiv:2501.14312}, 2025.

\bibitem[Chen et~al.(2024{\natexlab{a}})Chen, Deka, and Shi]{chen2024contributions}
Chen, Y., Deka, D., and Shi, Y.
\newblock Contributions of individual generators to nodal carbon emissions.
\newblock In \emph{Proceedings of the 15th ACM International Conference on Future and Sustainable Energy Systems}, pp.\  415--421, 2024{\natexlab{a}}.

\bibitem[Chen et~al.(2024{\natexlab{b}})Chen, Xie, Dang, Huang, Wu, and Jiao]{chen2023spatiotemporal}
Chen, Y., Xie, Y., Dang, X., Huang, B., Wu, C., and Jiao, D.
\newblock Spatiotemporal prediction of carbon emissions using a hybrid deep learning model considering temporal and spatial correlations.
\newblock \emph{Environmental Modelling \& Software}, 172:\penalty0 105937, 2024{\natexlab{b}}.
\newblock \doi{10.1016/j.envsoft.2023.105937}.

\bibitem[Chien et~al.(2022)Chien, Zhang, and Lin]{chien2022beyond}
Chien, A.~A., Zhang, C., and Lin, L.
\newblock Beyond pue: Flexible datacenters empowering the cloud to decarbonize.
\newblock \emph{USENIX Hot Carbon}, 2022.

\bibitem[Colangelo et~al.(2025)Colangelo, Coskun, Megrue, Roberts, Sengupta, Sivaram, Tiao, Vijaykar, Williams, Wilson, et~al.]{colangelo2025turning}
Colangelo, P., Coskun, A.~K., Megrue, J., Roberts, C., Sengupta, S., Sivaram, V., Tiao, E., Vijaykar, A., Williams, C., Wilson, D.~C., et~al.
\newblock Turning ai data centers into grid-interactive assets: Results from a field demonstration in phoenix, arizona.
\newblock \emph{arXiv preprint arXiv:2507.00909}, 2025.

\bibitem[Crankshaw et~al.(2017)Crankshaw, Wang, Zhou, Franklin, Gonzalez, and Stoica]{crankshaw2017clipper}
Crankshaw, D., Wang, X., Zhou, G., Franklin, M.~J., Gonzalez, J.~E., and Stoica, I.
\newblock Clipper: A $\{$Low-Latency$\}$ online prediction serving system.
\newblock In \emph{14th USENIX Symposium on Networked Systems Design and Implementation (NSDI 17)}, pp.\  613--627, 2017.

\bibitem[Duan et~al.(2024)Duan, Lu, Duanmu, Li, Zhang, Lin, Stoica, and Zhang]{duan2024muxserve}
Duan, J., Lu, R., Duanmu, H., Li, X., Zhang, X., Lin, D., Stoica, I., and Zhang, H.
\newblock Muxserve: Flexible spatial-temporal multiplexing for multiple llm serving.
\newblock \emph{arXiv preprint arXiv:2404.02015}, 2024.

\bibitem[Ferris \& Philpott(2025)Ferris and Philpott]{ferris2025optimizing}
Ferris, M.~C. and Philpott, A.~B.
\newblock Optimizing green energy systems.
\newblock \emph{INFORMS Journal on Optimization}, 2025.

\bibitem[Fu et~al.(2024)Fu, Zhu, Su, Qiao, Stoica, and Zhang]{fu2024efficient}
Fu, Y., Zhu, S., Su, R., Qiao, A., Stoica, I., and Zhang, H.
\newblock Efficient llm scheduling by learning to rank.
\newblock \emph{Advances in Neural Information Processing Systems}, 37:\penalty0 59006--59029, 2024.

\bibitem[{Gurobi Optimization, LLC}(2024)]{gurobi}
{Gurobi Optimization, LLC}.
\newblock {Gurobi Optimizer Reference Manual}, 2024.
\newblock URL \url{https://www.gurobi.com}.

\bibitem[Jain et~al.(1984)Jain, Chiu, Hawe, et~al.]{jain1984quantitative}
Jain, R.~K., Chiu, D.-M.~W., Hawe, W.~R., et~al.
\newblock A quantitative measure of fairness and discrimination.
\newblock \emph{Eastern Research Laboratory, Digital Equipment Corporation, Hudson, MA}, 21\penalty0 (1):\penalty0 2022--2023, 1984.

\bibitem[Jiang et~al.(2025)Jiang, Fu, Yao, He, Miao, Klimovic, Cui, Yuan, and Yoneki]{jiang2025demystifying}
Jiang, Y., Fu, F., Yao, X., He, G., Miao, X., Klimovic, A., Cui, B., Yuan, B., and Yoneki, E.
\newblock Demystifying cost-efficiency in llm serving over heterogeneous gpus.
\newblock \emph{arXiv preprint arXiv:2502.00722}, 2025.

\bibitem[Kakolyris et~al.(2024)Kakolyris, Masouros, Xydis, and Soudris]{kakolyris2024slo}
Kakolyris, A.~K., Masouros, D., Xydis, S., and Soudris, D.
\newblock Slo-aware gpu dvfs for energy-efficient llm inference serving.
\newblock \emph{IEEE Computer Architecture Letters}, 23\penalty0 (2):\penalty0 150--153, 2024.

\bibitem[Kakolyris et~al.(2025)Kakolyris, Masouros, Vavaroutsos, Xydis, and Soudris]{kakolyris2025throttll}
Kakolyris, A.~K., Masouros, D., Vavaroutsos, P., Xydis, S., and Soudris, D.
\newblock throttll’em: Predictive gpu throttling for energy efficient llm inference serving.
\newblock In \emph{2025 IEEE International Symposium on High Performance Computer Architecture (HPCA)}, pp.\  1363--1378. IEEE, 2025.

\bibitem[Kwon et~al.(2023)Kwon, Li, Zhuang, Sheng, Zheng, Yu, Gonzalez, Zhang, and Stoica]{kwon2023efficient}
Kwon, W., Li, Z., Zhuang, S., Sheng, Y., Zheng, L., Yu, C.~H., Gonzalez, J., Zhang, H., and Stoica, I.
\newblock Efficient memory management for large language model serving with pagedattention.
\newblock In \emph{Proceedings of the 29th symposium on operating systems principles}, pp.\  611--626, 2023.

\bibitem[Lacoste et~al.(2019)Lacoste, Luccioni, Schmidt, and Dandres]{lacoste2019quantifying}
Lacoste, A., Luccioni, A., Schmidt, V., and Dandres, T.
\newblock Quantifying the carbon emissions of machine learning.
\newblock \emph{arXiv preprint arXiv:1910.09700}, 2019.

\bibitem[Li et~al.(2023{\natexlab{a}})Li, Samsi, Gadepally, and Tiwari]{li2023clover}
Li, B., Samsi, S., Gadepally, V., and Tiwari, D.
\newblock Clover: Toward sustainable ai with carbon-aware machine learning inference service.
\newblock In \emph{Proceedings of the International Conference for High Performance Computing, Networking, Storage and Analysis}, pp.\  1--15, 2023{\natexlab{a}}.

\bibitem[Li et~al.(2020)Li, Wang, Zhang, Kontorinis, Kodakara, Lo, and Ranganathan]{li2020thunderbolt}
Li, S., Wang, X., Zhang, X., Kontorinis, V., Kodakara, S., Lo, D., and Ranganathan, P.
\newblock Thunderbolt:$\{$Throughput-Optimized$\}$,$\{$Quality-of-Service-Aware$\}$ power capping at scale.
\newblock In \emph{14th USENIX Symposium on Operating Systems Design and Implementation (OSDI 20)}, pp.\  1241--1255, 2020.

\bibitem[Li et~al.(2024)Li, Mughees, Chen, and Li]{li2024unseen}
Li, Y., Mughees, M., Chen, Y., and Li, Y.~R.
\newblock The unseen ai disruptions for power grids: Llm-induced transients.
\newblock \emph{arXiv preprint arXiv:2409.11416}, 2024.

\bibitem[Li et~al.(2025)Li, Hu, Choukse, Fonseca, Suh, and Gupta]{li2025ecoserve}
Li, Y., Hu, Z., Choukse, E., Fonseca, R., Suh, G.~E., and Gupta, U.
\newblock Ecoserve: Designing carbon-aware ai inference systems.
\newblock \emph{arXiv preprint arXiv:2502.05043}, 2025.

\bibitem[Li et~al.(2023{\natexlab{b}})Li, Zheng, Zhong, Liu, Sheng, Jin, Huang, Chen, Zhang, Gonzalez, et~al.]{li2023alpaserve}
Li, Z., Zheng, L., Zhong, Y., Liu, V., Sheng, Y., Jin, X., Huang, Y., Chen, Z., Zhang, H., Gonzalez, J.~E., et~al.
\newblock $\{$AlpaServe$\}$: Statistical multiplexing with model parallelism for deep learning serving.
\newblock In \emph{17th USENIX Symposium on Operating Systems Design and Implementation (OSDI 23)}, pp.\  663--679, 2023{\natexlab{b}}.

\bibitem[Lin \& Chien(2023)Lin and Chien]{lin2023adapting}
Lin, L. and Chien, A.~A.
\newblock Adapting datacenter capacity for greener datacenters and grid.
\newblock In \emph{Proceedings of the 14th ACM International Conference on Future Energy Systems}, pp.\  200--213, 2023.

\bibitem[Lindberg et~al.(2020)Lindberg, Lesieutre, and Roald]{lindberg2020environmental}
Lindberg, J., Lesieutre, B.~C., and Roald, L.
\newblock The environmental potential of hyper-scale data centers: Using locational marginal co $ \_2 $ emissions to guide geographical load shifting.
\newblock \emph{arXiv preprint arXiv:2010.03379}, 2020.

\bibitem[Lindberg et~al.(2022)Lindberg, Lesieutre, and Roald]{lindberg2022using}
Lindberg, J., Lesieutre, B.~C., and Roald, L.~A.
\newblock Using geographic load shifting to reduce carbon emissions.
\newblock \emph{Electric Power Systems Research}, 212:\penalty0 108586, 2022.

\bibitem[Lisa~Li et~al.(2025)Lisa~Li, Han, Suh, Delimitrou, Kazhamiaka, Choukse, Fonseca, Yu, Mace, and Gupta]{lisa2025fair}
Lisa~Li, Y., Han, L., Suh, G.~E., Delimitrou, C., Kazhamiaka, F., Choukse, E., Fonseca, R., Yu, L., Mace, J., and Gupta, U.
\newblock Fair, practical, and efficient carbon accounting for llm serving.
\newblock \emph{ACM SIGMETRICS Performance Evaluation Review}, 53\penalty0 (2):\penalty0 99--103, 2025.

\bibitem[Liu et~al.(2011)Liu, Lin, Wierman, Low, and Andrew]{liu2011greening}
Liu, Z., Lin, M., Wierman, A., Low, S.~H., and Andrew, L.~L.
\newblock Greening geographical load balancing.
\newblock \emph{ACM SIGMETRICS Performance Evaluation Review}, 39\penalty0 (1):\penalty0 193--204, 2011.

\bibitem[Low et~al.(2002)Low, Peterson, and Wang]{low2002understanding}
Low, S.~H., Peterson, L.~L., and Wang, L.
\newblock Understanding tcp vegas: a duality model.
\newblock \emph{Journal of the ACM (JACM)}, 49\penalty0 (2):\penalty0 207--235, 2002.

\bibitem[Luccioni et~al.(2024)Luccioni, Jernite, and Strubell]{luccioni2024power}
Luccioni, S., Jernite, Y., and Strubell, E.
\newblock Power hungry processing: Watts driving the cost of ai deployment?
\newblock In \emph{Proceedings of the 2024 ACM conference on fairness, accountability, and transparency}, pp.\  85--99, 2024.

\bibitem[Maji et~al.(2025)Maji, Hanafy, Wu, Irwin, Shenoy, and Sitaraman]{maji2025data}
Maji, D., Hanafy, W.~A., Wu, L., Irwin, D., Shenoy, P., and Sitaraman, R.~K.
\newblock Data centers carbon emissions at crossroads: An empirical study.
\newblock \emph{ACM SIGENERGY Energy Informatics Review}, 5\penalty0 (2):\penalty0 48--55, 2025.

\bibitem[Mao et~al.(2025)Mao, Xia, Wu, Chiang, Griggs, Bhardwaj, Yang, Shenker, and Stoica]{mao2025skyserve}
Mao, Z., Xia, T., Wu, Z., Chiang, W.-L., Griggs, T., Bhardwaj, R., Yang, Z., Shenker, S., and Stoica, I.
\newblock Skyserve: Serving ai models across regions and clouds with spot instances.
\newblock In \emph{Proceedings of the Twentieth European Conference on Computer Systems}, pp.\  159--175, 2025.

\bibitem[Mei et~al.(2024)Mei, Zhuang, Miao, Yang, Jia, and Vinayak]{mei2024helix}
Mei, Y., Zhuang, Y., Miao, X., Yang, J., Jia, Z., and Vinayak, R.
\newblock Helix: Distributed serving of large language models via max-flow on heterogeneous gpus.
\newblock \emph{Parameters}, 4\penalty0 (A100s):\penalty0 H100s, 2024.

\bibitem[Mo et~al.(1999)Mo, La, Anantharam, and Walrand]{mo1999analysis}
Mo, J., La, R.~J., Anantharam, V., and Walrand, J.
\newblock Analysis and comparison of tcp reno and vegas.
\newblock In \emph{IEEE INFOCOM'99. Conference on Computer Communications. Proceedings. Eighteenth Annual Joint Conference of the IEEE Computer and Communications Societies. The Future is Now (Cat. No. 99CH36320)}, volume~3, pp.\  1556--1563. IEEE, 1999.

\bibitem[Patel et~al.(2024{\natexlab{a}})Patel, Choukse, Zhang, Goiri, Warrier, Mahalingam, and Bianchini]{patel2024characterizing}
Patel, P., Choukse, E., Zhang, C., Goiri, {\'I}., Warrier, B., Mahalingam, N., and Bianchini, R.
\newblock Characterizing power management opportunities for llms in the cloud.
\newblock In \emph{Proceedings of the 29th ACM International Conference on Architectural Support for Programming Languages and Operating Systems, Volume 3}, pp.\  207--222, 2024{\natexlab{a}}.

\bibitem[Patel et~al.(2024{\natexlab{b}})Patel, Choukse, Zhang, Shah, Goiri, Maleki, and Bianchini]{patel2024splitwise}
Patel, P., Choukse, E., Zhang, C., Shah, A., Goiri, {\'I}., Maleki, S., and Bianchini, R.
\newblock Splitwise: Efficient generative llm inference using phase splitting.
\newblock In \emph{2024 ACM/IEEE 51st Annual International Symposium on Computer Architecture (ISCA)}, pp.\  118--132. IEEE, 2024{\natexlab{b}}.

\bibitem[PJM(2024)]{PJM2024LME}
PJM.
\newblock “five minute marginal emission rates".
\newblock \url{https://dataminer2.pjm.com/feed/fivemin_marginal_emissions/definition}, 2024.

\bibitem[{PJM Interconnection, L.L.C.}(2025)]{PJMDataMiner}
{PJM Interconnection, L.L.C.}
\newblock {Five Minute Marginal Emission Rates}.
\newblock \url{https://dataminer2.pjm.com/feed/fivemin_marginal_emissions}, 2025.
\newblock Data feed from PJM Data Miner 2. Accessed on September 23, 2025.

\bibitem[PuLP(2024)]{pulp}
PuLP, P.
\newblock Optimization with pulp, 2024.
\newblock URL \url{https://coin-or.github.io/pulp/}.

\bibitem[Qiu et~al.(2024{\natexlab{a}})Qiu, Mao, Patke, Cui, Jha, Wang, Franke, Kalbarczyk, Ba{\c{s}}ar, and Iyer]{qiu2024power}
Qiu, H., Mao, W., Patke, A., Cui, S., Jha, S., Wang, C., Franke, H., Kalbarczyk, Z., Ba{\c{s}}ar, T., and Iyer, R.~K.
\newblock Power-aware deep learning model serving with $\{$$\mu$-Serve$\}$.
\newblock In \emph{2024 USENIX Annual Technical Conference (USENIX ATC 24)}, pp.\  75--93, 2024{\natexlab{a}}.

\bibitem[Qiu et~al.(2024{\natexlab{b}})Qiu, Mao, Patke, Cui, Jha, Wang, Franke, Kalbarczyk, Ba{\c{s}}ar, and Iyer]{qiu2024efficient}
Qiu, H., Mao, W., Patke, A., Cui, S., Jha, S., Wang, C., Franke, H., Kalbarczyk, Z.~T., Ba{\c{s}}ar, T., and Iyer, R.~K.
\newblock Efficient interactive llm serving with proxy model-based sequence length prediction.
\newblock \emph{arXiv preprint arXiv:2404.08509}, 2024{\natexlab{b}}.

\bibitem[Radovanovi{\'c} et~al.(2022)]{radovanovic2022carbon}
Radovanovi{\'c}, A. et~al.
\newblock Carbon-aware computing for datacenters.
\newblock \emph{IEEE Transactions on Power Systems}, 38\penalty0 (2):\penalty0 1270--1280, 2022.

\bibitem[Review(2025)]{MIT2025AI}
Review, M.~T.
\newblock “ai and our energy future".
\newblock \url{https://www.technologyreview.com/supertopic/ai-energy-package/}, 2025.

\bibitem[Samsi et~al.(2023)Samsi, Zhao, McDonald, Li, Michaleas, Jones, Bergeron, Kepner, Tiwari, and Gadepally]{samsi2023words}
Samsi, S., Zhao, D., McDonald, J., Li, B., Michaleas, A., Jones, M., Bergeron, W., Kepner, J., Tiwari, D., and Gadepally, V.
\newblock From words to watts: Benchmarking the energy costs of large language model inference.
\newblock In \emph{2023 IEEE High Performance Extreme Computing Conference (HPEC)}, pp.\  1--9. IEEE, 2023.

\bibitem[Sheng et~al.(2024)Sheng, Cao, Li, Zhu, Li, Zhuo, Gonzalez, and Stoica]{sheng2024fairness}
Sheng, Y., Cao, S., Li, D., Zhu, B., Li, Z., Zhuo, D., Gonzalez, J.~E., and Stoica, I.
\newblock Fairness in serving large language models.
\newblock In \emph{18th USENIX Symposium on Operating Systems Design and Implementation (OSDI 24)}, pp.\  965--988, 2024.

\bibitem[Srivatsa et~al.(2025)Srivatsa, He, Abhyankar, Li, and Zhang]{srivatsapreble}
Srivatsa, V., He, Z., Abhyankar, R., Li, D., and Zhang, Y.
\newblock Preble: Efficient distributed prompt scheduling for llm serving.
\newblock In \emph{The Thirteenth International Conference on Learning Representations}, 2025.

\bibitem[Stojkovic et~al.(2025{\natexlab{a}})Stojkovic, Zhang, Goiri, Choukse, Qiu, Fonseca, Torrellas, and Bianchini]{stojkovic2025tapas}
Stojkovic, J., Zhang, C., Goiri, {\'I}., Choukse, E., Qiu, H., Fonseca, R., Torrellas, J., and Bianchini, R.
\newblock Tapas: Thermal-and power-aware scheduling for llm inference in cloud platforms.
\newblock In \emph{Proceedings of the 30th ACM International Conference on Architectural Support for Programming Languages and Operating Systems, Volume 2}, pp.\  1266--1281, 2025{\natexlab{a}}.

\bibitem[Stojkovic et~al.(2025{\natexlab{b}})Stojkovic, Zhang, Goiri, Torrellas, and Choukse]{stojkovic2025dynamollm}
Stojkovic, J., Zhang, C., Goiri, {\'I}., Torrellas, J., and Choukse, E.
\newblock Dynamollm: Designing llm inference clusters for performance and energy efficiency.
\newblock In \emph{2025 IEEE International Symposium on High Performance Computer Architecture (HPCA)}, pp.\  1348--1362. IEEE, 2025{\natexlab{b}}.

\bibitem[Sukprasert et~al.(2024)Sukprasert, Souza, Bashir, Irwin, and Shenoy]{sukprasert2024limitations}
Sukprasert, T., Souza, A., Bashir, N., Irwin, D., and Shenoy, P.
\newblock On the limitations of carbon-aware temporal and spatial workload shifting in the cloud.
\newblock In \emph{Proceedings of the Nineteenth European Conference on Computer Systems}, pp.\  924--941, 2024.

\bibitem[Tang et~al.(2019)Tang, Wang, Wang, and Chu]{tang2019impact}
Tang, Z., Wang, Y., Wang, Q., and Chu, X.
\newblock The impact of gpu dvfs on the energy and performance of deep learning: An empirical study.
\newblock In \emph{Proceedings of the Tenth ACM International Conference on Future Energy Systems}, pp.\  315--325, 2019.

\bibitem[Wang et~al.(2018)Wang, Wang, Gao, Zhang, Chen, Ng, and Ooi]{wang2018rafiki}
Wang, W., Wang, S., Gao, J., Zhang, M., Chen, G., Ng, T.~K., and Ooi, B.~C.
\newblock Rafiki: Machine learning as an analytics service system.
\newblock \emph{arXiv preprint arXiv:1804.06087}, 2018.

\bibitem[Wang et~al.(2006)Wang, Palaniswami, and Low]{wang2006application}
Wang, W.-H., Palaniswami, M., and Low, S.~H.
\newblock Application-oriented flow control: fundamentals, algorithms and fairness.
\newblock \emph{IEEE/ACM Transactions On Networking}, 14\penalty0 (6):\penalty0 1282--1291, 2006.

\bibitem[Wang et~al.(2014)Wang, Wang, Miller, McElmurry, Miller, and Rogers]{wang2014locational}
Wang, Y., Wang, C., Miller, C., McElmurry, S., Miller, S., and Rogers, M.
\newblock Locational marginal emissions: Analysis of pollutant emission reduction through spatial management of load distribution.
\newblock \emph{Applied energy}, 119:\penalty0 141--150, 2014.

\bibitem[Wierman et~al.(2014)Wierman, Liu, Liu, and Mohsenian-Rad]{wierman2014opportunities}
Wierman, A., Liu, Z., Liu, I., and Mohsenian-Rad, H.
\newblock Opportunities and challenges for data center demand response.
\newblock In \emph{International Green Computing Conference}, pp.\  1--10. IEEE, 2014.

\bibitem[Wright(1997)]{wright1997primal}
Wright, S.~J.
\newblock \emph{Primal-dual interior-point methods}.
\newblock SIAM, 1997.

\bibitem[Wu et~al.(2023)Wu, Zhong, Zhang, Liu, Liu, Sun, Huang, Liu, and Jin]{wu2023fast}
Wu, B., Zhong, Y., Zhang, Z., Liu, S., Liu, F., Sun, Y., Huang, G., Liu, X., and Jin, X.
\newblock Fast distributed inference serving for large language models.
\newblock \emph{arXiv preprint arXiv:2305.05920}, 2023.

\bibitem[Wu et~al.(2014)Wu, Chang, and Chan]{wu2014green}
Wu, C.-M., Chang, R.-S., and Chan, H.-Y.
\newblock A green energy-efficient scheduling algorithm using the dvfs technique for cloud datacenters.
\newblock \emph{Future Generation Computer Systems}, 37:\penalty0 141--147, 2014.

\bibitem[Ye et~al.(2025)Ye, Chen, Lai, Lin, Zhang, Wang, Chen, Kasikci, Grover, Krishnamurthy, et~al.]{ye2025flashinfer}
Ye, Z., Chen, L., Lai, R., Lin, W., Zhang, Y., Wang, S., Chen, T., Kasikci, B., Grover, V., Krishnamurthy, A., et~al.
\newblock Flashinfer: Efficient and customizable attention engine for llm inference serving.
\newblock \emph{arXiv preprint arXiv:2501.01005}, 2025.

\bibitem[Yu et~al.(2025)Yu, Taneja, Lin, and Zhang]{yu2025voltanallm}
Yu, J., Taneja, A., Lin, J., and Zhang, M.
\newblock Voltanallm: Feedback-driven frequency control and state-space routing for energy-efficient llm serving.
\newblock \emph{arXiv preprint arXiv:2509.04827}, 2025.

\bibitem[Zheng et~al.(2023{\natexlab{a}})Zheng, Chiang, Sheng, Li, Zhuang, Wu, Zhuang, Li, Lin, Xing, Gonzalez, Stoica, and Zhang]{zheng2023lmsyschat1m}
Zheng, L., Chiang, W.-L., Sheng, Y., Li, T., Zhuang, S., Wu, Z., Zhuang, Y., Li, Z., Lin, Z., Xing, E.~P., Gonzalez, J.~E., Stoica, I., and Zhang, H.
\newblock Lmsys-chat-1m: A large-scale real-world llm conversation dataset, 2023{\natexlab{a}}.

\bibitem[Zheng et~al.(2023{\natexlab{b}})Zheng, Yin, Xie, Huang, Sun, Yu, Cao, Kozyrakis, Stoica, Gonzalez, et~al.]{zheng2023efficiently}
Zheng, L., Yin, L., Xie, Z., Huang, J., Sun, C., Yu, C., Cao, S., Kozyrakis, C., Stoica, I., Gonzalez, J.~E., et~al.
\newblock Efficiently programming large language models using sglang.
\newblock 2023{\natexlab{b}}.

\bibitem[Zhong et~al.(2024)Zhong, Liu, Chen, Hu, Zhu, Liu, Jin, and Zhang]{zhong2024distserve}
Zhong, Y., Liu, S., Chen, J., Hu, J., Zhu, Y., Liu, X., Jin, X., and Zhang, H.
\newblock $\{$DistServe$\}$: Disaggregating prefill and decoding for goodput-optimized large language model serving.
\newblock In \emph{18th USENIX Symposium on Operating Systems Design and Implementation (OSDI 24)}, pp.\  193--210, 2024.

\end{thebibliography}
\bibliographystyle{mlsys2025}
\appendix
\newpage
\section*{}
\newpage
\section{Pool-Level Details}
\label{Pool-level Simulations}
 \begin{figure*}[t] 
      \centering
      \includegraphics[width=1\linewidth]{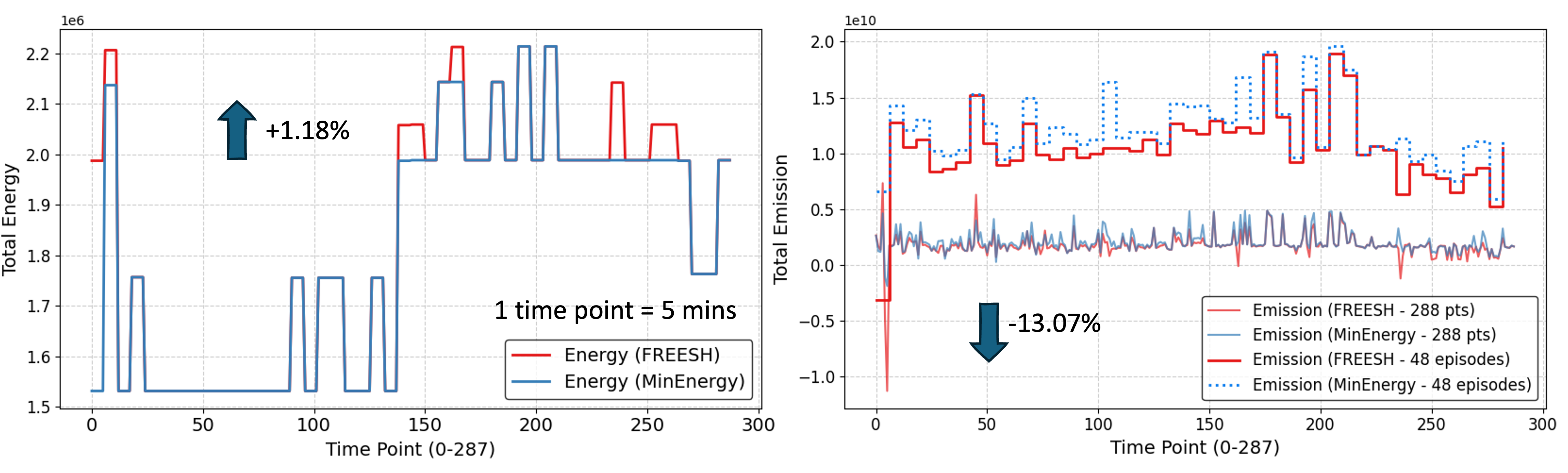}
      \caption{Overall energy consumption and carbon emission across 1-day pool-level optimization.}\label{fig:T1_performance}
    \end{figure*}
   
 \begin{figure*} [t]
      \centering 
      \includegraphics[width=1\linewidth]{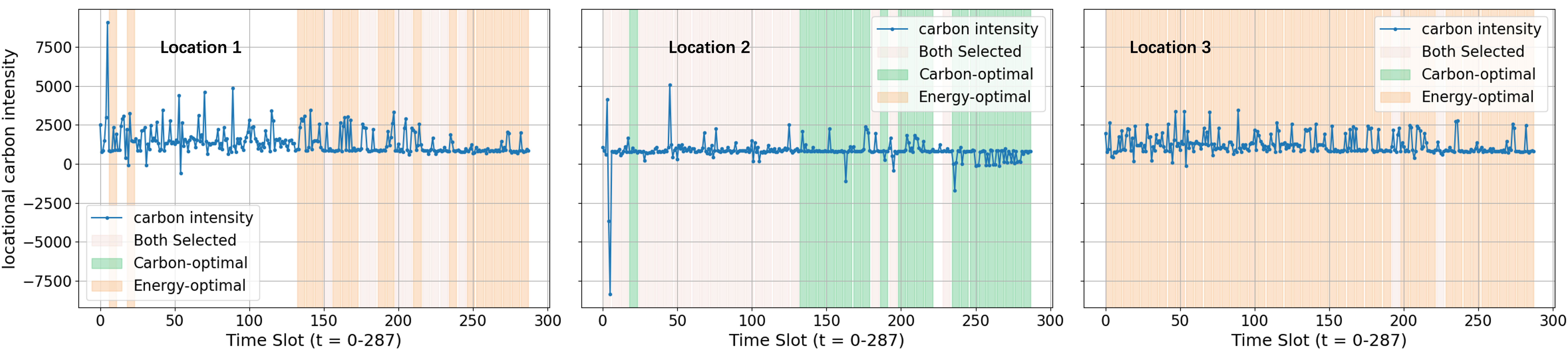}
      \vspace{-1em}
      \caption{Location variation response to the spatiotemporal carbon intensity.} \label{carbon_intensity}
    \end{figure*}
    
\subsection{Overall Energy Optimal VS Emission Optimal}
The pool-level optimization in FREESH aims for emission optimality, which it typically equates with energy optimality, assuming this sufficiently reduces carbon emissions. However, our simulations on a 24-hour, open-source production trace demonstrate a critical limitation of this approach. By explicitly modeling spatiotemporal variations in carbon intensity, our FREESH (Eq. \ref{Upper-IP}) achieves a 13.07\% reduction in carbon emissions at the cost of a minor 1.18\% increase in energy consumption compared to \texttt{MinEnergy} as demonstrated in Fig. \ref{fig:T1_performance}. The \texttt{MinEnergy} formulation only minimizes the total energy used for serving LLM without considering emission intensity associated with electricity.  This finding underscores that a purely energy-optimal approach to LLM serving can be myopic, failing to achieve a global carbon optimum by overlooking crucial geographical and temporal factors associated with locational emission rates.


\subsection{Carbon Signal Tracing} From Fig. \ref{fig:workloadflowofFREESH}, we can see that the IP model's location selections are time-varying, with different decisions made for energy-optimal and emission-optimal operations. Location 2 is consistently chosen due to its sufficient GPU capacity. However, FREESH can still identify and select the period with the lowest carbon intensity, such as during time slots 230-240. While during timeslots 0-4 as shown in Table \ref{tab:wrapped-table}, FREESH can adjust the request partition to deploy more workers at location 2 to trace the negative carbon intensity, which brings 207 \% emission reduction. This demonstrates that FREESH effectively tracks carbon intensity and identifies the most carbon-efficient location for LLM services. 

\begin{table}[htb]
  \hspace{-5em}
  \vspace{-1em}
  \caption{Partition Decision within timeslots 0-4 in Fig. \ref{carbon_intensity}}\label{Partition_Decision}
\setlength{\tabcolsep}{0.001pt}
\begin{tabular}{c|cc}
\hline
Set                  & \multicolumn{1}{c}{\texttt{w/o Part}: \{\textbf{P. 1}\}~~$\rightarrow$} &FREESH: \{P. 1,~\textbf{P. 2}\} \\ \hline
Emission                    & 0.41 kg                                                             & -0.44 kg ($\downarrow 207\%$)                             \\ \hline\cline{1-1}
\multirow{2}{*}{Allocation} & \textbf{loc2}: SS-m1, SL-m1,                                                        & \textbf{loc1}: SL-m4, \textbf{loc2}:SS-m1,                       \\
                            & LL-m1, \textbf{loc3}:LL-m2                                                          & LS-m2*2, l3:LL-m3                              \\ \hline
\end{tabular}
  \label{tab:wrapped-table}
\end{table}

\subsection{Request Partition}
In this paper, we apply the following two discrete partition ways listed in Table \ref{Patition}, and the pool-level optimization (Eq. \ref{Upper-IP}) optimizes for the optimal request partition set. In practice, any feasible partitions could be applied in FREESH.

Specifically, \textbf{Partition 1} is determined empirically: we first exclude the most extreme 2.5\% of samples from both ends of the request distribution, and then use the median token length within the remaining 95\% range as the threshold for distinguishing short and long requests. This approach reflects an observation-based heuristic division that captures the majority of typical workloads.  
In contrast, \textbf{Partition 2} adopts a simpler, statistically grounded method using the 50th percentile (median) of token length as the boundary, representing a balanced partition between short and long requests.

\begin{table}[h]
  \centering
  {\fontsize{8}{10}\selectfont
\setlength{\tabcolsep}{7pt}
  \caption{Request classification according to token length.}
\label{Patition}
\begin{tabular}{c|cccc}
\hline
            & \multicolumn{2}{c}{Input} & \multicolumn{2}{c}{Output} \\ \hline
Type        & Short (S)    & Long (L)   & Short (S)    & Long (L)    \\ \hline
Partition 1 & {[}0, 184)        & {[}184, 4096{]}    & {[}0, 444)        & {[}444, 4096{]}     \\
Partition 2 & {[}0, 25)        & {[}25, 4096{]}    & {[}0, 232)        & {[}232, 4096{]}     \\ \hline
\end{tabular}}
\end{table}

\section{GPU-Level Details} \label{GPU-Level Simulations}

\subsection{Theoretical Analysis for MIAD}
\label{sec:MIAD}
In our frequency scaling algorithm, we dynamically adjust each TP instance's frequency to maximize utility received by each request $i$ with received service rate $y_i$, minimize the power consumption $P(f)$, while subject to the GPU instance's performance limit $r(f)$. Though we only take $f$ as the control knob in this algorithm, it well generalizes to scenarios where multiple queries are considered with distributed serving rate $y_i$. Here we present the proof of Proposition 1 as follows,
\begin{proof}
The Lagrangian  of \eqref{equ:utility_max} is as follows:
\begin{subequations}
\begin{align}\label{lower_level_dual}
L(\boldsymbol{y},f,r) = \sum_i{U_i}(y_i)-\beta{P(f)}+\gamma(r(f)-\sum_i{y_i}).
\end{align}
\end{subequations}

Given strictly concave and continuously differentiable utility functions $U_i(y_i),\; i=1,...,N$ and convex, continuously differentiable $P(f)$, \eqref{equ:utility_max} is a concave optimization problem. The KKT conditions for original utility maximization problem~\eqref{equ:utility_max} are as follows:
\begin{subequations}  
\begin{align}
\label{equ:lagrangian}
&\frac{\partial{L}}{\partial{y}_{i}} = U'_i(x_i)-\gamma = 0, U'_i({y^{*}_{i}}) = \gamma, y^{*}_{i}= {U'}^{-1}_{i}(\gamma) \\
&\frac{\partial{L}}{\partial{f}} = -\beta P'(f) + \gamma{r'(f^{*})} =0, P'(f^*) = \frac{\gamma}{\beta}r'(f^{*}), \\
&\sum_i{y}_i \leq r(f),\\
&y_i \geq 0, \\
&\underline{f} \leq f \leq \overline{f}, \\
&\gamma \geq 0, 
\\
&\gamma(r(f)-\sum_i(y_i))=0.
\end{align}
\end{subequations}
Based on standard duality theory~\cite{boyd2004convex}, if there exists $f$ and $y$ strictly satisfying the constraint,  there is zero duality gap, thus primal global optimum can be recovered from the dual optimum. Following the primal-dual update rule \eqref{equ:primal-dual}, it is guaranteed to converge to the global optimum~\cite{wright1997primal}.
\end{proof}

\begin{figure} [t]
      \centering 
      \includegraphics[width=1\linewidth]{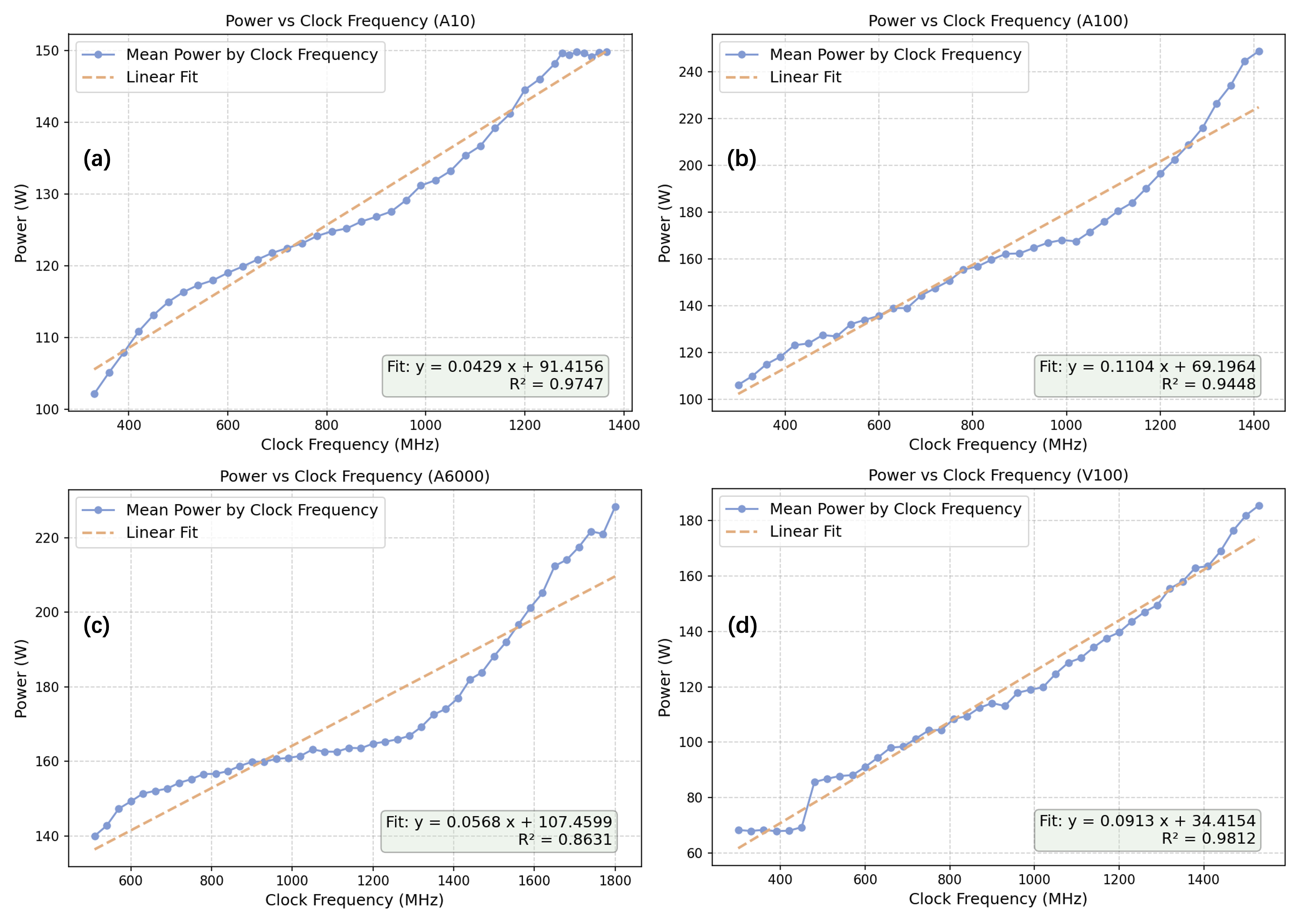}
      \vspace{-1em}
      \caption{Frequency-Power curve across GPU types} \label{fig:gpu-comparison}
    \end{figure}

In the practical implementation, we favor lowered frequency as long as SLO can be met to reduce power consumption. However, when ${\gamma}r'(f) \ge \beta{P}'(f)$, we increase $f$, which can be concluded as the dynamic GPU frequency scaling principle \eqref{eq:miad_alg}. Also in practice, the assumption that $U_i(x_i)$ is a concave function and $P(f)$ is a convex function are mild, considering candidate utility function for log-utility, and the fitted linear frequency-power curve shown in Fig. \ref{fig:gpu-comparison}.
    

        
    

\subsection{Frequency and Power Variation in MIAD}
\begin{figure*}[t!]
  \centering

  \begin{minipage}{0.9\textwidth}
    \centering
    \includegraphics[width=0.4\linewidth]{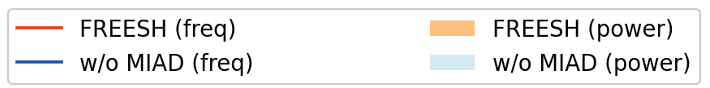}

    \begin{minipage}{0.49\linewidth}
      \centering
      \subcaptionbox{SS (short--short)\label{fig:miad_a}}{%
        \includegraphics[width=\linewidth]{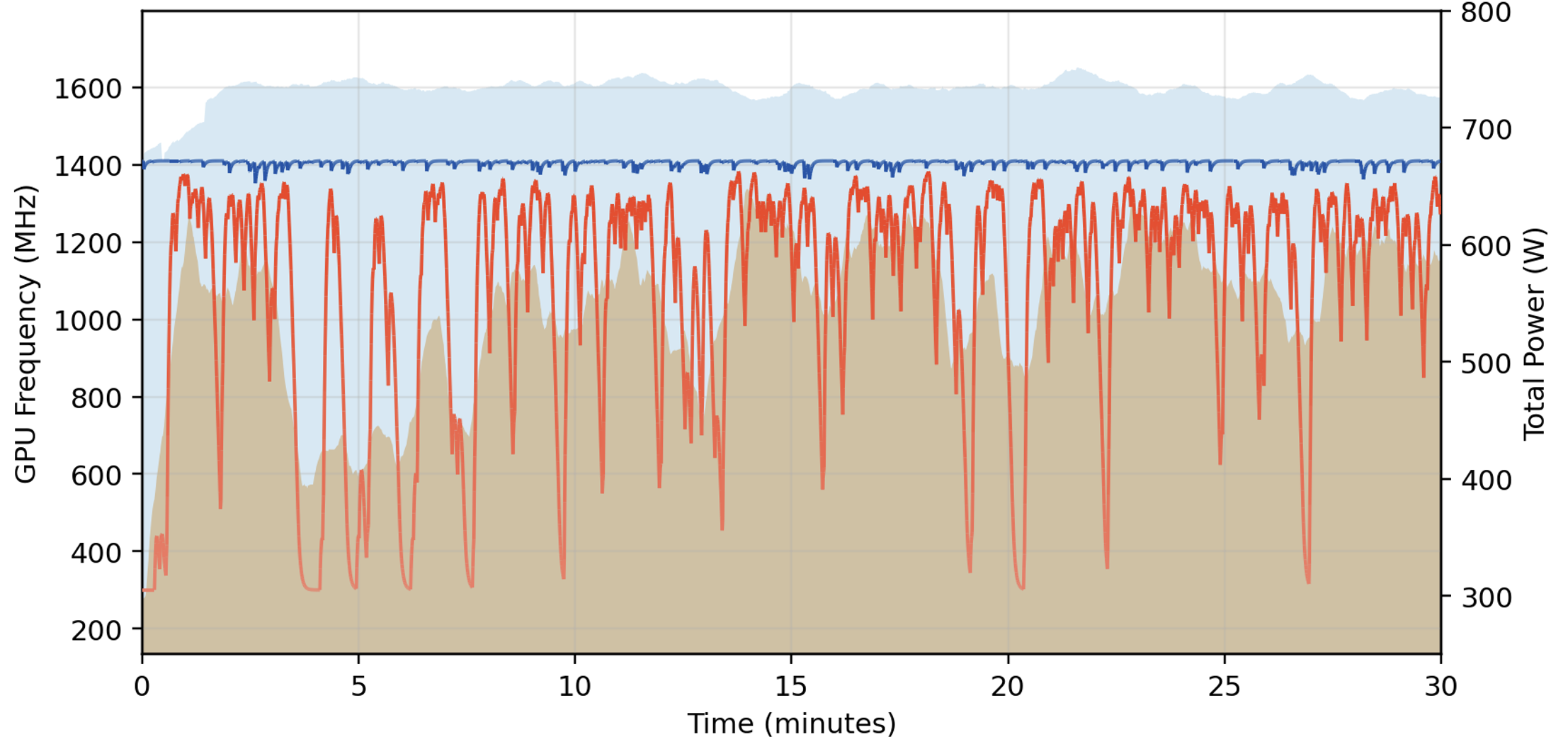}}
    \end{minipage}\hfill
    \begin{minipage}{0.49\linewidth}
      \centering
      \subcaptionbox{SL (short--long)\label{fig:miad_b}}{%
        \includegraphics[width=\linewidth]{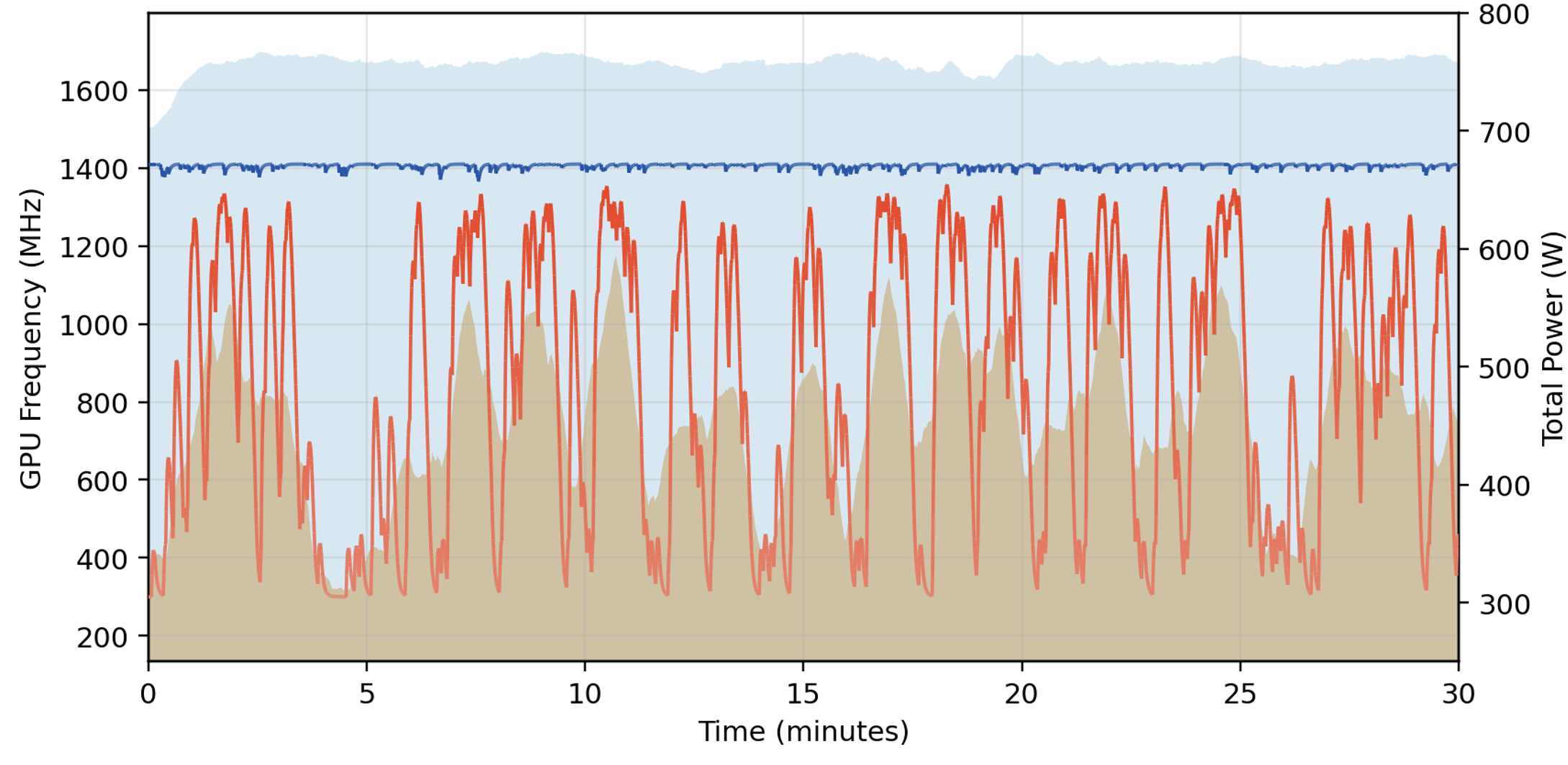}}
    \end{minipage}

    \vspace{0.9\baselineskip}

    \begin{minipage}{0.49\linewidth}
      \centering
      \subcaptionbox{LS (long--short)\label{fig:miad_c}}{%
        \includegraphics[width=\linewidth]{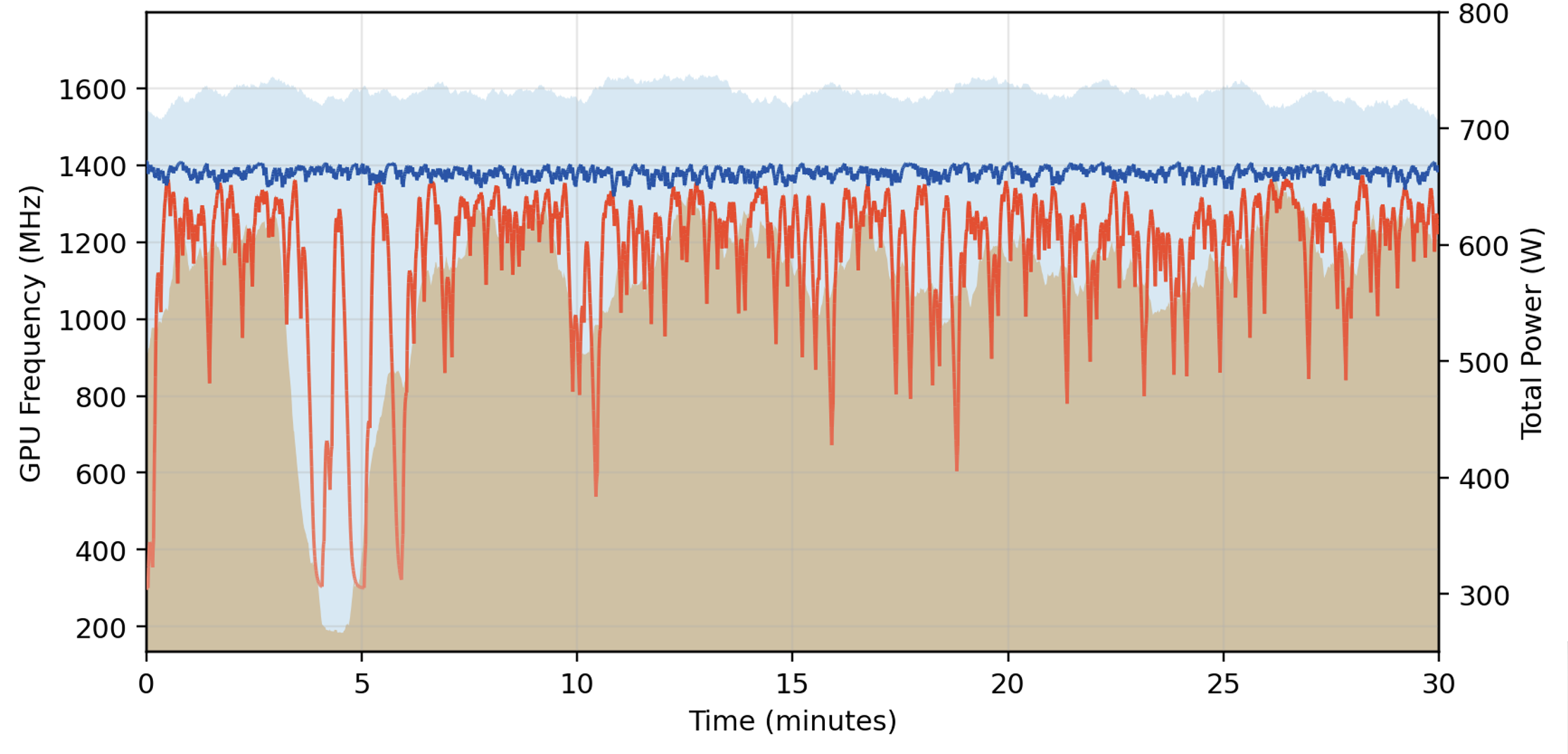}}
    \end{minipage}\hfill
    \begin{minipage}{0.49\linewidth}
      \centering
      \subcaptionbox{LL (long--long)\label{fig:miad_d}}{%
        \includegraphics[width=\linewidth]{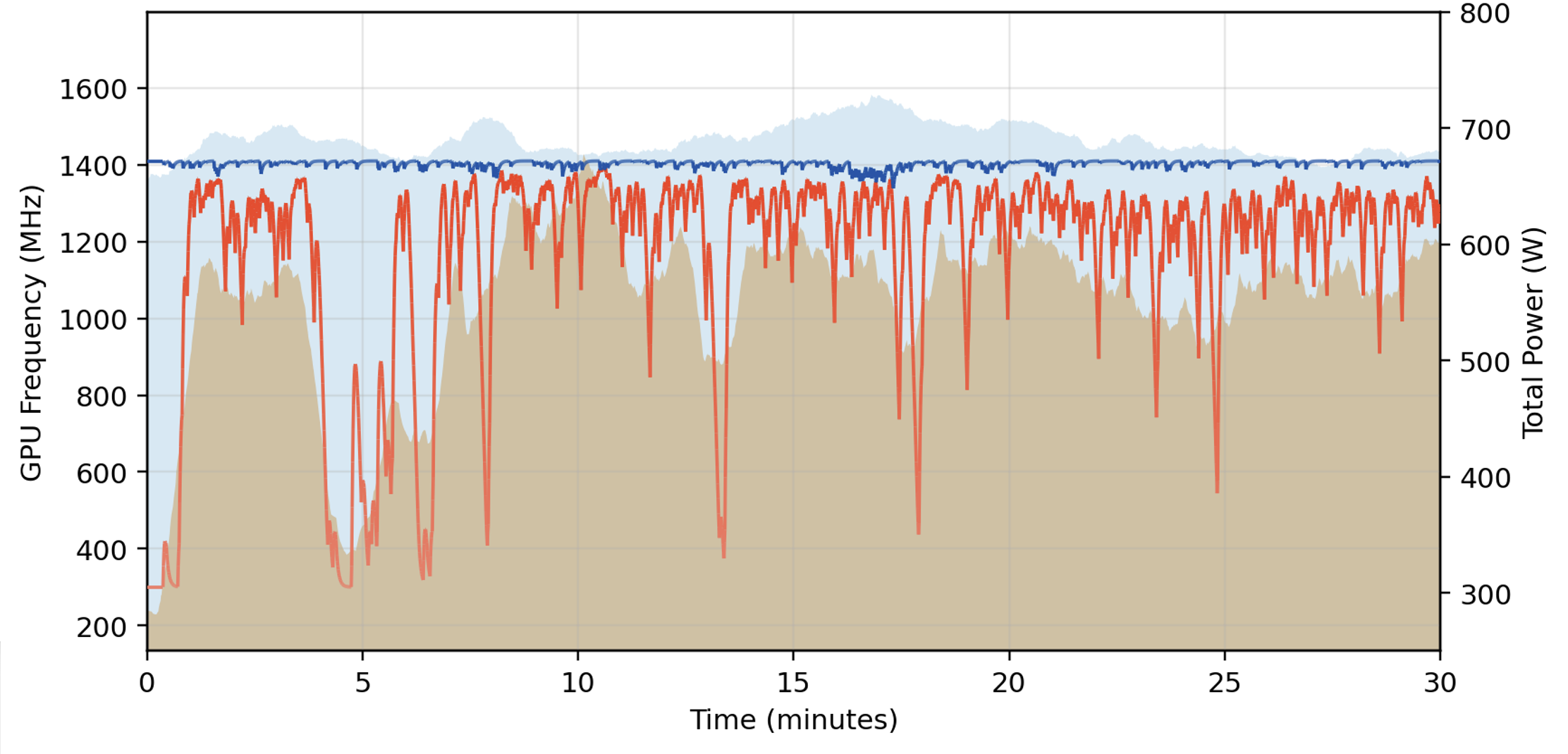}}
    \end{minipage}
  \end{minipage}

  \caption{%
    Frequency and power comparison of the MIAD algorithm across four request types.
    The comparison shown is for Llama3-70B, using data from workers running on 2x A100 80GB SXM GPUs.
    Subplots (a), (b), (c), and (d) show the GPU frequency and power consumption for FREESH (with MIAD) versus the baseline (\texttt{w/o MIAD}) under
    SS (short--short), SL (short--long), LS (long--short), and LL (long--long) request types, respectively.
  }
  \label{fig:miad_freq_power_comparison}
\end{figure*}

In practice, we note that setting the GPU frequency (e.g., via nvidia-smi) incurs non-negligible overheads. 
Figure \ref{fig:miad_freq_power_comparison} provides a comprehensive comparison of system performance with and without the MIAD frequency scaling algorithm across all four request types. 
As shown in all four subplots, the baseline (\texttt{w/o MIAD}) maintains a static, high GPU frequency, consistently operating near the A100's peak of 1410 MHz. 
This static high-frequency operation results in a persistently high total power consumption for the dual-card setup, which remains stable at over 700W, irrespective of the actual workload.

In sharp contrast, FREESH, powered by the MIAD algorithm, exhibits highly dynamic behavior. 
The GPU frequency is actively adjusted based on real-time load, oscillating between the minimum floor of approximately 300 MHz during idle periods and the peak of 1410 MHz when high performance is required. 
Consequently, the power consumption also dynamically scales, fluctuating between approximately 200W and 650W. 
This demonstrates MIAD's core benefit: it achieves significant energy savings when the load is low, while still scaling up to provide maximum performance when the load surges.

    

\begin{figure*}[h!]
    \centering
    \resizebox{0.8\textwidth}{!}{%
        \begin{minipage}{\textwidth}
            \centering
            \subfloat[Idle period with a short busy burst.]{%
                \includegraphics[width=0.48\textwidth]{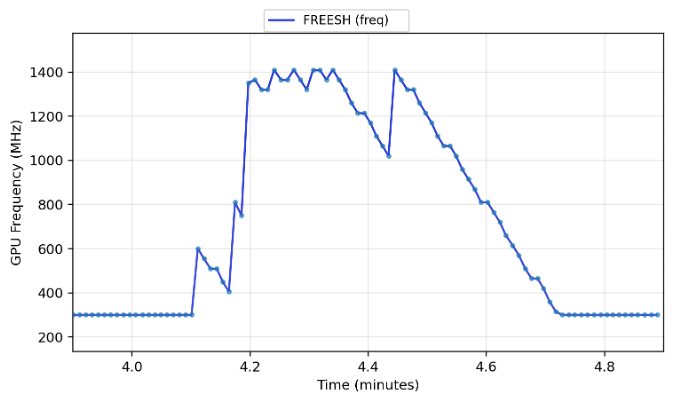}
            }
            \hfill
            \subfloat[Busy period with a short idle segment.]{%
                \includegraphics[width=0.48\textwidth]{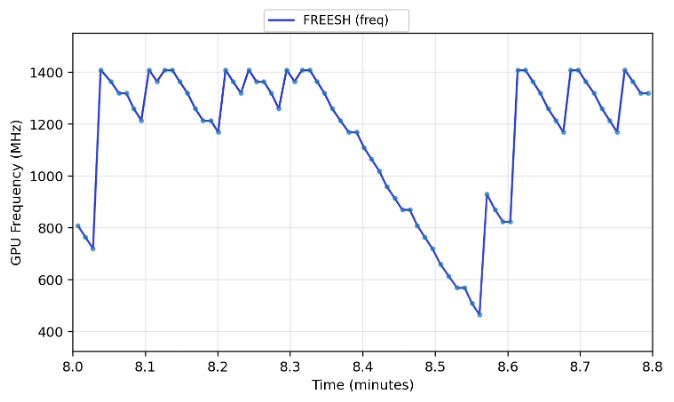}
            }\\[4pt]
            \subfloat[Continuously busy interval.]{%
                \includegraphics[width=0.48\textwidth]{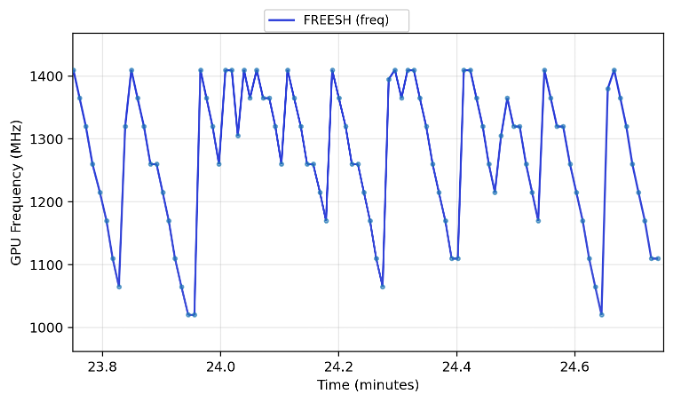}
            }
            \hfill
            \subfloat[Long idle period with sporadic requests.]{%
                \includegraphics[width=0.48\textwidth]{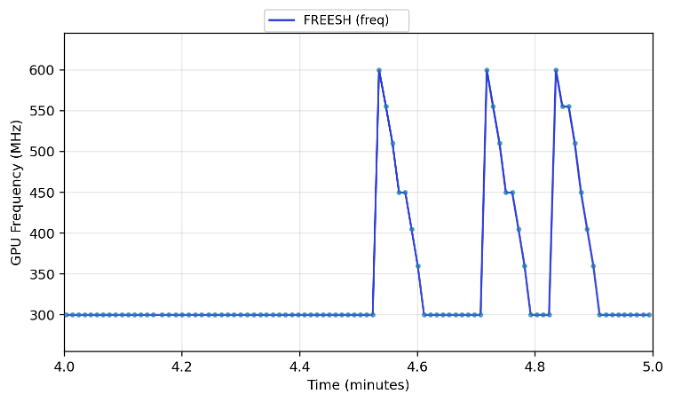}
            }
        \end{minipage}
    }
    \caption{
        A zoomed-in view of the MIAD frequency scaling mechanism, corresponding to the 4th--5th minute interval of Figure~\ref{fig:miad_freq_power_comparison}(a), under four typical traffic patterns:
        (a) mostly idle $\rightarrow$ short busy burst,
        (b) mostly busy $\rightarrow$ short idle period,
        (c) continuously busy,
        and (d) continuously idle with sporadic requests.
        These cases together highlight how MIAD applies multiplicative increase to react immediately to SLO violations or rising load, and additive decrease to steadily return to the energy-efficient operating point.
    }
    \label{fig:miad_zoom_detail}
\end{figure*}

The precise mechanics of this adjustment are illustrated in Figure~\ref{fig:miad_zoom_detail}, which now presents four fine-grained, one-minute snapshots of the MIAD algorithm in action. 
These zoomed-in views jointly demonstrate the two core principles of MIAD: Multiplicative Increase (MI) and Additive Decrease (AD), and, more importantly, they show that MIAD can closely track rapid changes in traffic.

Specifically, Figure~\ref{fig:miad_zoom_detail}(a) shows a mostly idle period where a short burst of requests suddenly arrives. 
When the burst starts, MIAD immediately performs a multiplicative increase to raise the GPU frequency to a high level, and then, after the burst subsides, it slowly and additively decreases the frequency back to the minimum, reflecting the MI–AD policy.

Figure~\ref{fig:miad_zoom_detail}(b) captures the opposite case: the system is under a relatively busy period, but there is a short idle window. 
MIAD senses the reduced load and gradually reduces the frequency in an additive, stepwise manner, while still keeping the system ready to ramp up again once the load returns.

Figure~\ref{fig:miad_zoom_detail}(c) depicts a continuously busy interval. 
In this case, MIAD keeps the frequency at a high level with only small oscillations, showing that the controller can sustain performance when the workload remains high.

Finally, Figure~\ref{fig:miad_zoom_detail}(d) illustrates a long, very-idle period. 
Here the frequency stays near the minimum most of the time, and only when sporadic requests arrive does MIAD issue short, spike-like multiplicative increases; after each request is served, the frequency quickly falls back to the minimum. 
Taken together, these four scenarios show that MIAD can effectively sense and track traffic dynamics.

\section{Request-Level Details}

\subsection{LLF Implementation}

To facilitate reproducibility and further analysis, this Section presents the complete implementation of the LLF scheduling algorithm.  
The algorithm has been introduced in the main text as the core LLM scheduling strategy.  
Algorithm~\ref{alg:llf} below illustrates the detailed LLF scheduling procedure within the simulation framework.

\begin{algorithm}[H]
\caption{Least Laxity First Scheduling Scheme}
\label{alg:llf}
\begin{algorithmic}[1]
\STATE \textbf{Inputs:} set of requests $\{i\}$ with $(t_{i,\mathrm{arrival}}, T_{i,\mathrm{remain}}, W_i)$; tick size $\Delta t$
\FOR{$t = 0, \Delta t, 2\Delta t, \dots$ until all requests finish}
  \STATE Let $\mathcal{A}(t) = \{ i \mid t_{i,\mathrm{arrival}} \le t,\ T_{i,\mathrm{remain}} > 0 \}$ \COMMENT{active requests}
  \FOR{each $i \in \mathcal{A}(t)$}
    \STATE update laxity $\ell_i(t)$ using \eqref{eq:lax}
  \ENDFOR
  \STATE Sort $\mathcal{A}(t)$ by increasing $\ell_i(t)$
  \STATE Assign available workers to requests with smallest $\ell_i(t)$
  \STATE Execute each selected request for $\Delta t$
  \STATE Update remaining time $T_{i,\mathrm{remain}} \leftarrow T_{i,\mathrm{remain}} - \Delta t$
\ENDFOR
\end{algorithmic}
\end{algorithm}

\subsection{Fairness Metric for Requests} 

To evaluate the fairness achieved under the LLF scheduling policy, requests of the same workload type are divided into $K$ disjoint client groups according to their output token counts using quartile-based partitioning. Let $K=4$ denote the number of clients, and let $C_k$ denote the set of requests belonging to client $k$.
For each request $i \in C_k$, a fairness evaluation window $W_i$ is defined consistently with the scheduling window as $W_i = \alpha_{LLF} \times Lat$, where $\alpha_{LLF} = 1.4$ in our experiments.  
Let $\mathrm{finish}_{i,t}$ denote the completion time of its $t$-th token. 
The number of tokens that finish within the fairness window is given by
\[
x_i = \sum_t \mathbf{1}\{\mathrm{finish}_{i,t} - t_{i,\text{arrival}} \leq W_i\},
\]
where $\mathbf{1}\{\cdot\}$ denotes the indicator function, which equals 1 if the condition inside the braces is true and 0 otherwise. 
The total number of timely-served tokens for client $k$ is 
\[
X_k = \sum_{i \in C_k} x_i.
\]

Finally, the fairness metric across all clients is quantified using the Jain Fairness Index:
\begin{equation}
\mathrm{Jain\,Fairness} = \frac{\left(\sum_{k=1}^{K} X_k\right)^2}{K \times \sum_{k=1}^{K} X_k^2},
\end{equation}
which ranges from $0$ (completely unfair) to $1$ (perfectly fair). This metric measures whether each client receives a comparable share of timely-served tokens within its fairness window.


The excellent fairness of the preemptive LLF scheduling algorithm stems from its dynamic priority mechanism based on \textbf{laxity}, a metric that integrates a request's remaining workload with its deadline urgency. This design prevents long tasks from being continuously starved by short ones, as the priority of a long task naturally increases as its deadline approaches. As a result, the scheduler effectively balances the service experience across different clients by proactively reallocating computational resources to the most time-sensitive tasks, thereby avoiding systemic bias towards any single class of requests.

In addition to fairness, the timeliness of LLF scheduling is assessed through the server-side SLO violation rate, defined as follows.
A server-side SLO violation occurs if a request finishes after the violation window
$window_i =\beta_ {LLF} \times Lat$ with $\beta_{LLF}{=}5$; the empirical violation rate is the fraction of requests violating this condition:
\begin{equation}
\label{eq:violation}
\mathrm{Violation\,Rate} \;=\; \frac{1}{N}\sum_{i=1}^{N}\mathbf{1}\!\left\{\,t_{i, \text{finish}}-t_{i, \text{arrival}} > window_i,\right\}.
\end{equation}

where $t_{i,\mathrm{finish}}$ denote the completion time of request $i$.


\subsection{The Calculation of Toy Example}

Figure~\ref{fig:toyexample} illustrates three requests, $R_0$, $R_1$, and $R_2$, which arrive sequentially at $t=0$, $t=1$, and $t=2$, respectively. Their output lengths are 10, 2, and 1 tokens, while the system throughput is fixed at 1~token/s. The scheduling policy updates every second, meaning that only one token can be produced per time slot. The average latency is measured as the completion time minus the arrival time per token.

The metrics reported in Figure~\ref{fig:toyexample} are defined as follows.

\textbf{Latency (s/token).}
For each request $R_i$, the average per-token latency is defined as
\[
    \text{Latency}(R_i) = \frac{t_{i, \text{finish}} - t_{i, \text{arrival}}}{N_{\text{tokens}}(R_i)
}.
\]
This per-token normalization enables comparison across requests with different output lengths.

For the three scheduling policies---FCFS, SRTF, and LLF---the token generation timelines follow the sequences shown in Figure~\ref{fig:toyexample}. By substituting each request’s arrival time, first-token time, and completion time into the above two formulas, the corresponding latency and TTFT values presented in the figure can be directly obtained.

For example, under LLF scheduling, the latency of $R_1$ is calculated as follows: 
it finishes at the 4th second and arrives at the 1st second, so $\text{Latency}(R_1) = \frac{4 - 1}{2} = 1.5$.

\subsection{Performance Comparison}

The experiment uses the four scheduling algorithms (FCFS, SRTF, EDF, LLF). Under the standard configuration with 8 GPUs, 100 requests are sent to the LLM with a concurrency level of 20.  

\begin{figure*}[t]
  \centering
  \includegraphics[width=1\linewidth]{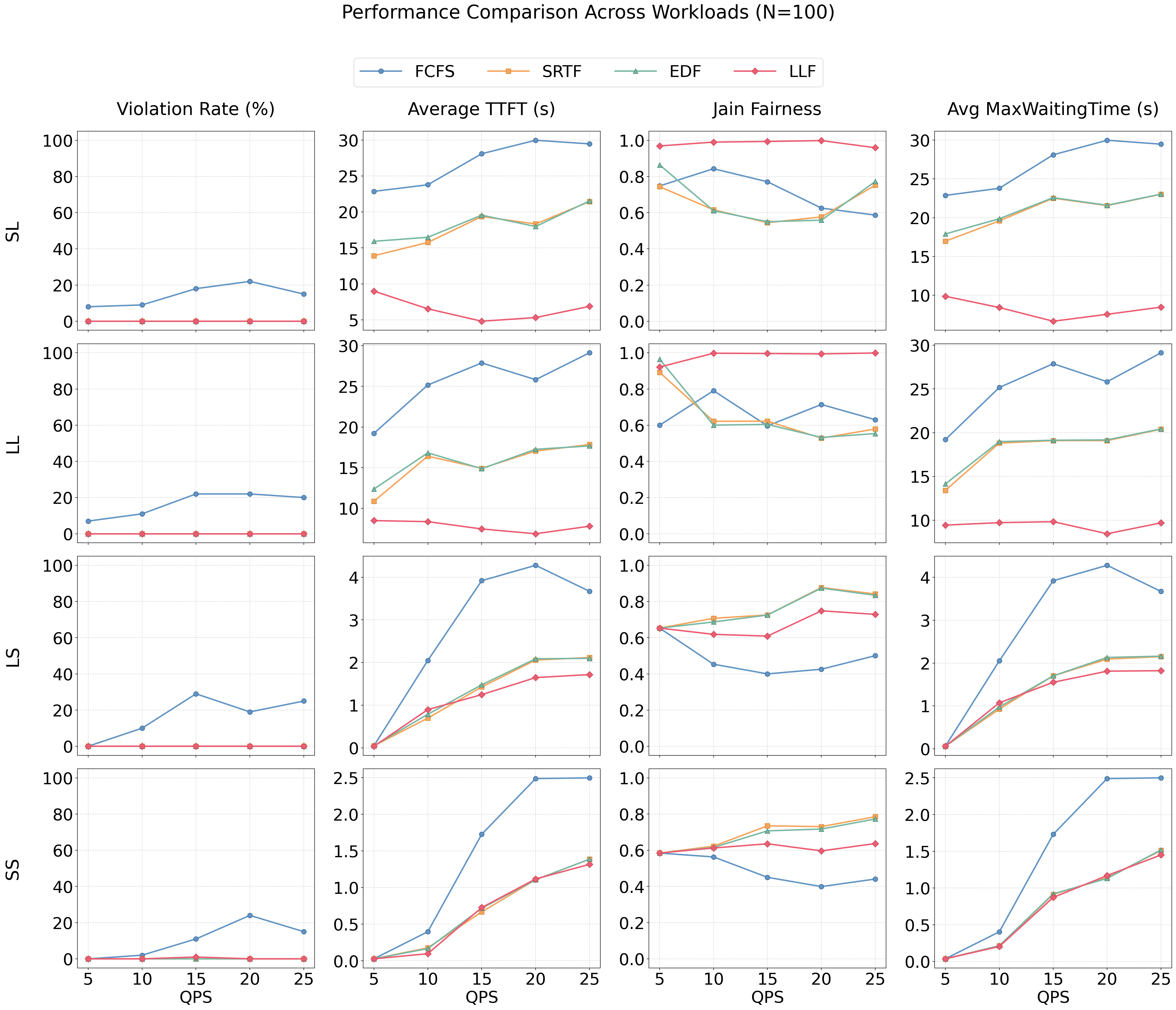}
  \caption{Different metrics under varying QPS rate.}
    \label{fig:qps-scan}
\end{figure*}

The experiment sweeps the QPS (Queries per second) rate across $\{5, 10, 15, 20, 25\}$ queries/s.  
For each QPS rate, four metrics are recorded: violation rate (Violation\%), average time-to-first-token (Avg TTFT), fairness (Jain Fairness), and average maximum waiting time (Avg MaxWaitingTime).  
Avg MaxWaitingTime is defined as the average, over all requests, of each request’s maximum waiting duration, where the per-request value is the larger of (i) the delay from arrival to the first token and (ii) the longest interval between two consecutive token completions.As shown in Figure~\ref{fig:qps-scan} and Table~\ref{tab:performance_summary}, the curves of the four algorithms under different workloads follow the same set of metrics and allow a direct visual comparison.

Across all four algorithms (FCFS, SRTF, EDF, LLF), LLF consistently achieves a superior latency–fairness balance.  
The results show the following patterns:

\textbf{(1) Long-output workloads.}  
LLF exhibits clear advantages: its Avg TTFT and Avg MaxWaitingTime are significantly lower than those of other algorithms, indicating that LLF effectively prioritizes time-critical tasks under heavy load, alleviating queue buildup and improving responsiveness.  
In contrast, FCFS tends to experience head-of-line blocking as QPS increases, leading to higher violation rates.

\textbf{(2) Short-output workloads.}  
When responses are short, all algorithms achieve similar Avg TTFT and Avg MaxWaitingTime, as short tasks dominate total latency and scheduling has limited influence.

\textbf{(3) Fairness.}  
In short-output scenarios, Jain Fairness values across algorithms are nearly identical.  
However, in long-output scenarios, LLF attains noticeably higher fairness, reflecting a more balanced allocation of compute resources.

\begin{table*}[thb]
\centering
\caption{The sensitivity of queuing method's SLO Performance to workload level with varying request types.}
\label{tab:performance_summary}
{\fontsize{8}{10}\selectfont
\setlength{\tabcolsep}{4pt}
\begin{tabular}{llccc>{\columncolor[HTML]{F0F0F0}}c
ccc>{\columncolor[HTML]{F0F0F0}}c
ccc>{\columncolor[HTML]{F0F0F0}}c
ccc>{\columncolor[HTML]{F0F0F0}}c}
\toprule
\multicolumn{2}{l}{} & \multicolumn{4}{c}{\textbf{SS}} & \multicolumn{4}{c}{\textbf{SL}} & \multicolumn{4}{c}{\textbf{LS}} & \multicolumn{4}{c}{\textbf{LL}} \\
\cmidrule(lr){3-6} \cmidrule(lr){7-10} \cmidrule(lr){11-14} \cmidrule(lr){15-18}
\multicolumn{2}{l}{} & EDF & SRTF & FCFS & LLF & EDF & SRTF & FCFS & LLF & EDF & SRTF & FCFS & LLF & EDF & SRTF & FCFS & LLF \\
\midrule
\multirow{4}{*}{QPS=5} & Violation \% $ (\downarrow)$ & \textbf{0.0} & \textbf{0.0} & \textbf{0.0} & \textbf{0.0} & \textbf{0.0} & \textbf{0.0} & 8.0 & \textbf{0.0} & \textbf{0.0} & \textbf{0.0} & \textbf{0.0} & \textbf{0.0} & \textbf{0.0} & \textbf{0.0} & 7.0 & \textbf{0.0} \\
& TTFT (s)$ (\downarrow)$ & \textbf{0.0} & \textbf{0.0} & \textbf{0.0} & \textbf{0.0} & 15.9 & 13.9 & 22.9 & \textbf{9.0} & 0.1 & 0.1 & 0.1 & \textbf{0.0} & 12.4 & 10.9 & 19.2 & \textbf{8.5} \\
& Jain Fair.$ (\uparrow)$& \textbf{0.6} & \textbf{0.6} & \textbf{0.6} & \textbf{0.6} & 0.9 & 0.7 & 0.7 & \textbf{1.0} & \textbf{0.7} & \textbf{0.7} & \textbf{0.7} & \textbf{0.7} & \textbf{1.0} & 0.9 & 0.6 & 0.9 \\
& MaxWaiting (s) $ (\downarrow)$  & \textbf{0.0} & \textbf{0.0} & \textbf{0.0} & \textbf{0.0} & 17.9 & 17.0 & 22.9 & \textbf{9.9} & \textbf{0.1} & \textbf{0.1} & \textbf{0.1} & \textbf{0.1} & 14.1 & 13.4 & 19.2 & \textbf{9.4} \\
\midrule
\multirow{4}{*}{QPS=15} & Violation \% $ (\downarrow)$ & \textbf{0.0} & \textbf{0.0} & 11.0 & 1.0 & \textbf{0.0} & \textbf{0.0} & 18.0 & \textbf{0.0} & \textbf{0.0} & \textbf{0.0} & 29.0 & 3.0 & \textbf{0.0} & \textbf{0.0} & 22.0 & \textbf{0.0} \\
& TTFT (s) $ (\downarrow)$  & 0.7 & \textbf{0.7} & 1.7 & 0.7 & 19.6 & 19.4 & 28.1 & \textbf{4.8} & 1.5 & 1.4 & 3.9 & \textbf{1.3} & 14.9 & 14.9 & 27.9 & \textbf{7.5} \\
& Jain Fair.$ (\uparrow)$ & 0.7 & \textbf{0.7} & 0.4 & 0.6 & 0.5 & 0.5 & 0.8 & \textbf{1.0} & 0.7 & \textbf{0.7} & 0.4 & 0.6 & 0.6 & 0.6 & 0.6 & \textbf{1.0} \\
& MaxWaiting (s) $ (\downarrow)$  & 0.9 & 0.9 & 1.7 & \textbf{0.9} & 22.6 & 22.5 & 28.1 & \textbf{6.6} & \textbf{1.7} & \textbf{1.7} & 3.9 & 1.6 & 19.1 & \textbf{19.1} & 27.9 & \textbf{9.8} \\
\midrule
\multirow{4}{*}{QPS=25} & Violation \% $ (\downarrow)$  & \textbf{0.0} & \textbf{0.0} & 16.0 & \textbf{0.0} & \textbf{0.0} & \textbf{0.0} & 15.0 & \textbf{0.0} & \textbf{0.0} & \textbf{0.0} & 28.0 & 1.0 & \textbf{0.0} & \textbf{0.0} & 20.0 & \textbf{0.0} \\
& TTFT (s) $ (\downarrow)$  & 1.4 & 1.4 & 2.5 & \textbf{1.3} & 21.6 & 21.5 & 29.5 & \textbf{6.9} & 2.1 & 2.1 & 3.7 & \textbf{1.7} & 17.7 & 17.8 & 29.1 & \textbf{7.8} \\
& Jain Fair.$ (\uparrow)$  & 0.8 & \textbf{0.8} & 0.4 & 0.6 & \textbf{0.8} & 0.8 & 0.6 & \textbf{1.0} & 0.8 & \textbf{0.8} & 0.5 & 0.7 & 0.6 & 0.6 & 0.6 & \textbf{1.0} \\
& MaxWaiting (s) $ (\downarrow)$  & 1.5 & \textbf{1.5} & 2.5 & 1.5 & 23.0 & 23.0 & 29.5 & \textbf{8.5} & 2.2 & \textbf{2.2} & 3.7 & 1.8 & 20.4 & \textbf{20.4} & 29.1 & \textbf{9.7} \\
\bottomrule
\end{tabular}}
\end{table*}
 
\subsection{Time Window Selection}

In the simulation environment, two key parameters are introduced: $\alpha_{LLF}$ and $\beta_{LLF}$. 
The parameter $\alpha_{LLF}$ controls the fairness evaluation window (representing the client-side latency tolerance), 
while $\beta_{LLF}$ controls the violation window (representing the operator-side latency tolerance). 

\begin{figure*}[htbp]
  \centering
  \includegraphics[width=1\linewidth]{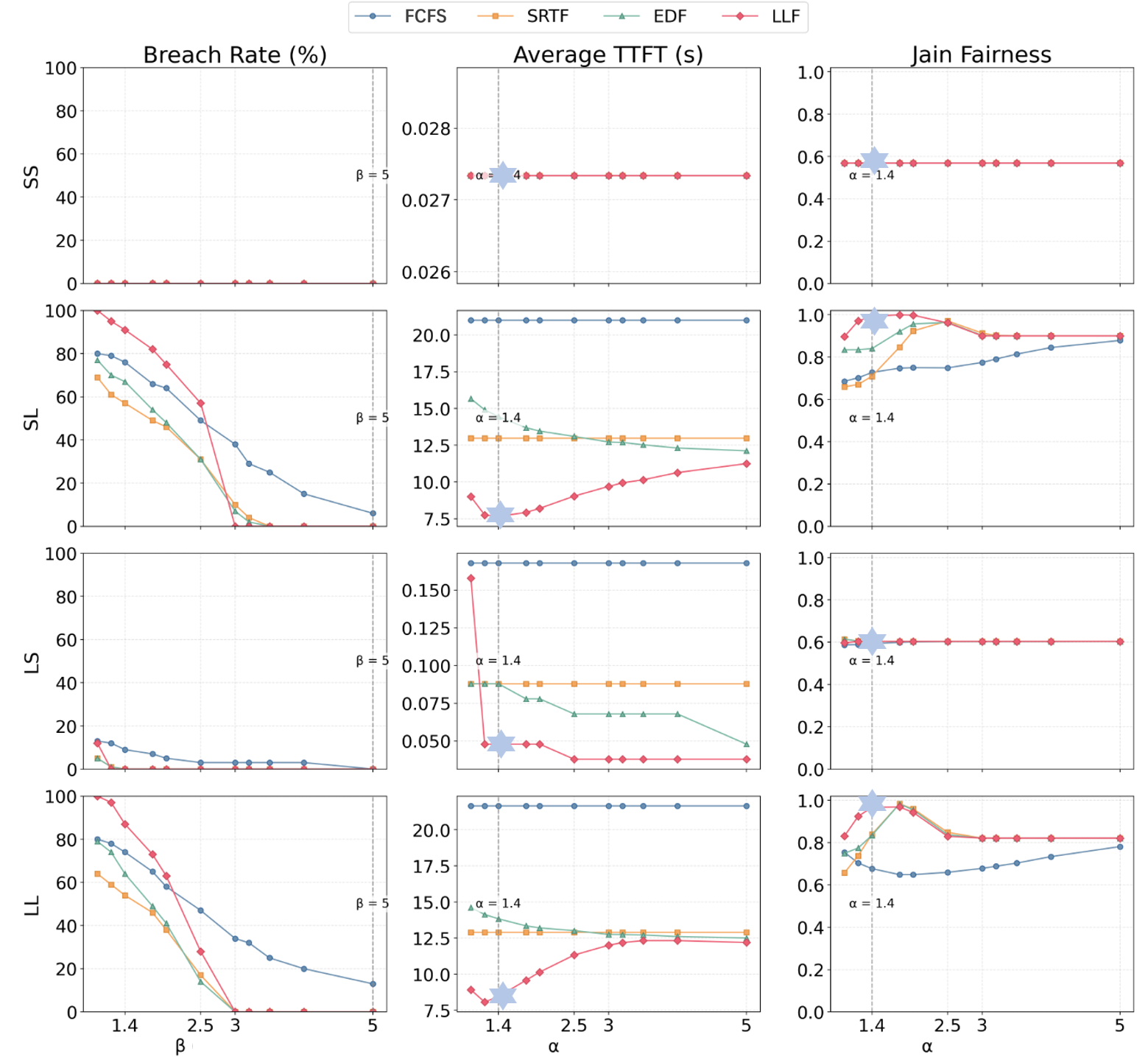}
  \caption{Comparison of $\alpha, \; \beta$ parameters on LLF performance with varying request type (SS, SL, LS, LL). The selection of $\alpha_{LLF} = 1.4$ and $\beta_{LLF} = 5$ achieves the optimal performance.}
    \label{fig:slo}
\end{figure*}

Following the configuration used in DynamoLLM~\cite{stojkovic2025dynamollm}, $\beta_{LLF}$ is fixed at $5$, 
representing the upper-bound latency window tolerated by the system operator under the worst-case condition. 
However, clients typically expect response times within a much smaller time window than this upper bound. 
Therefore, for the client-side fairness window, we conduct experiments with $\alpha_{LLF} \in [1.4,\,1.6]$. 
The results show that within this range, LLF achieves the best balance between fairness and latency, 
and thus $\alpha_{LLF}$ is finally fixed at $1.4$ as the default experimental setting.
As shown in Figure~\ref{fig:slo}, when $\alpha_{LLF}$ is selected as 1.4, both Average TTFT and Jain Fairness perform well.

\section{Implementation Details}
\subsection{Time Overheads}

We evaluated the time overheads introduced by the components of the FREESH controller to validate its efficiency in real-time decision-making. A detailed breakdown of these overheads is presented in Table \ref{tab:time-overheads}.

\begin{table}[h]
\centering
\caption{Time Overheads of the FREESH Controller.}
\label{tab:time-overheads}
\begin{adjustbox}{width=0.48\textwidth}
\begin{tabular}{@{}llc@{}}
\toprule
\textbf{Component} & \textbf{Frequency} & \textbf{Avg. Time} \\ 
\midrule
\multicolumn{3}{l}{\textit{\textbf{1. Slow-Timescale Overheads}}} \\
Traffic Forecast (LSTM) & Every 30 mins & 45.2 ms \\
Emission Forecast & Every 5 mins & $<$ 100 ms \\
MILP Solving  & Every 30 mins & $<$ 60 s\\
TP Switching (Actuation) & On config change & 35.4 ms \\
\midrule
\multicolumn{3}{l}{\textit{\textbf{2. Fast-Timescale Overheads}}} \\
Request Classification (BERT) & Per request & 16 ms \\
LLF Calculation & Per schedule tick & 5 ms \\
Frequency Control (MIAD) & Every 1 s & $<$ 20 ms \\
\bottomrule
\end{tabular}
\end{adjustbox}
\end{table}

It is important to emphasize that all latency measurements were obtained under our specific experimental configuration and system scale. These results are notably affected by both the hardware configuration of the controller node (e.g., CPU model) and the scale of the optimization problem, as illustrated in Figure \ref{solu_tra}.

For slow-timescale (pool-level) overheads, the primary cost arises from the optimization solver. Depending on the problem scale, the solver’s runtime ranged from 0.08 s to 19.36 s. In the table, we conservatively report this as $< 60$ s. Since this computation is executed only once every 30 minutes (1800 s) and runs asynchronously, its impact on real-time serving is negligible.

For fast-timescale (request-level) overheads, all components operate on the millisecond scale, including request classification (16 ms), LLF queue calculation (5 ms), and frequency control ($< 20$ ms). These real-time processing costs are negligible compared to the end-to-end LLM inference latency, which typically ranges from a few seconds to tens of seconds, confirming that the FREESH framework operates efficiently without introducing noticeable performance bottlenecks.

\subsection{SLO Thresholds}

As referenced in Section D.7, the MIAD frequency scaling mechanism relies on a set of latency-based thresholds to detect real-time congestion. These thresholds, detailed in Table~\ref{tab:slo_thresholds}, are defined per request type to trigger the multiplicative frequency increase of the MIAD controller.

\begin{table}[htbp]
  \centering
  \caption{SLO Thresholds for MIAD Congestion Control.}
  \label{tab:slo_thresholds}
  \resizebox{0.85\linewidth}{!}{%
    \begin{tabular}{@{}cccc@{}}
      \toprule
      \textbf{Request Type} & \textbf{TBT (ms)} & \textbf{TTFT (ms)} & \textbf{E2E Latency (s)} \\
      \midrule
      SS & 150 & 300 & 7 \\
      SL & 150 & 300 & 27 \\
      LS & 150 & 600 & 10 \\
      LL & 150 & 600 & 30 \\
      \bottomrule
    \end{tabular}%
  }
\end{table}

The thresholds are calibrated based on the distinct performance characteristics of each request class:
\begin{itemize}
    \item \textbf{Time-Between-Token (TBT):} A uniform threshold of 150~ms is applied across all request types. This metric is primarily sensitive to the token generation (decode) phase, which has similar performance expectations regardless of the initial prompt.
    
    \item \textbf{TTFT:} The TTFT threshold is dependent on the input prompt length. Requests with short inputs (SS and SL) use a tighter threshold of 300~ms, while requests with long inputs (LS and LL) are given a more lenient 600~ms. This accounts for the additional processing time (prefill phase) required for longer prompts before token generation can begin.
    
    \item \textbf{End-to-End (E2E) Latency:} The E2E thresholds (7s, 27s, 10s, 30s) are derived from the average end-to-end latencies empirically measured during real GPU experiments for each respective request type. These measured averages serve as the practical baselines for the E2E latency thresholds, acting as a final backstop to detect significant overall request slowdown.
\end{itemize}

\subsection{LLM Traffic Forecast}
In FREESH, the traffic prediction module is designed to forecast the incoming traffic into four distinct service request types: SS, SL, LS, and LL. To achieve this objective, we have developed a prediction framework based on the Long-Short-Term Memory (LSTM) model. The methodology consists of the following stages:

For model training, we selected a public Azure dataset and extracted one week of complete conversation logs as our training data. First, we processed the data into a time series format: by establishing a fundamental time window of 5 seconds, we counted the number of incoming requests for each type within that window to form a single data point.

We used an LSTM network to capture the temporal dependencies of the request traffic. The model takes a sequence of continuous data points from a past period as input and predicts the request arrivals at each time point within the next 30 minutes.

The model outputs the forecast results for the subsequent 30 minutes (which correspond to 360 data points of 5-second intervals). The 30-minute horizon is then divided into 6 consecutive 5-minute intervals, and we calculate the maximum QPS for each interval. The specific calculation involves identifying the 5-second data point with the highest request volume within each 5-minute interval and dividing its request count by 5 to obtain the peak QPS for that interval. Ultimately, we use this peak QPS as the core predictive indicator for the traffic intensity of each request type over the next half-hour.

\subsection{Emission Forecast}
Accurately forecasting the real-time carbon emissions of geographically distributed Large Language Model (LLM) serving clusters is critical for achieving sustainable computing and enabling dynamic load balancing. This task is highly challenging as it constitutes a complex \textbf{spatiotemporal} forecasting problem: emissions are not only dynamic over time but are also influenced by the spatial location of different data center nodes (e.g., local grid carbon intensity) and their interactions.\cite{chen2023spatiotemporal}

We employ a hybrid deep learning model that integrates a Gated Recurrent Unit (GRU) and a Graph Convolutional Network (GCN)~\cite{chen2023spatiotemporal}. When applied to LLM cluster forecasting, time-series data from each data center, such as GPU utilization and request load, is first fed into a GRU network to capture its intrinsic temporal evolution. Subsequently, all data centers are structured as a geographical graph, upon which the GCN operates to learn and integrate the spatial correlations between nodes, such as how the low carbon intensity of one node might influence global workload distribution and total emissions.

This spatiotemporal hybrid model achieved a coefficient of determination (R²) exceeding \textbf{0.95} in its original monthly data predictions, significantly outperforming models that consider only a single dimension. Its architecture is well-suited for high-frequency, dynamic data, thereby providing an advanced and validated technical framework for achieving high-precision, spatiotemporal \textbf{five-minute-level} carbon emission predictions for LLM clusters.

\subsection{Request Classifier}\label{sec:prediction}
To support efficient request scheduling in our system, we train a request classifier that jointly (i) predicts the expected output length of a request (measured in tokens) and (ii) produces a binary label indicating \emph{short} vs.\ \emph{long} responses based on the predicted length. Training data are drawn from the \textsc{LMSYS-Chat-1M} dataset \cite{zheng2023lmsyschat1m}, where we take the first turn of user--assistant interaction prompts at temperature~0 and submit them to \textsc{Llama-3-70B}. We collect 50K request--response pairs, each containing the user prompt, the model reply, and the corresponding token counts for both prompt and answer.

We find that a lightweight model can predict the LLM type well given the input query. Building on the insight of \citet{fu2024efficient} that pretrained encoders such as BERT can be adapted to predict response lengths, we design our classifier to accommodate requests with a maximum length of 4,096 tokens in our dataset. Since XLM-RoBERTa can process at most 512 tokens per input, directly encoding such long contexts is infeasible. To address this, we partition each request into 512-token chunks, encode each chunk with XLM-RoBERTa, and aggregate the chunk embeddings using a bidirectional LSTM. The aggregated representation is then passed to a regression head that predicts \(\log(\text{output\_tokens}+1)\); we exponentiate at inference to recover the expected output length. Finally, we derive the binary label by thresholding the predicted length at 444 tokens, a cutoff selected from the empirical distribution of response sizes in our corpus.

\begin{table}[!htbp]
\centering
\caption{Configuration and hyperparameters of request classifier.}
\label{tab:req-config}
\setlength{\tabcolsep}{0.8pt}\begin{tabular}{ll}
\toprule
Component / Parameter & Setting \\
\midrule
Encoder backbone & XLM-RoBERTa (base) \\
Max request length & 4096 tokens \\
Chunk size & 512 tokens \\
Aggregator & BiLSTM (Hidden size = 384) \\
Output head & Linear regressor on aggregated state \\
Long/short threshold & 444 tokens \\
Optimizer & AdamW (Learning rate $2\times 10^{-5}$) \\
Batch size & 16 \\
Epochs & 5 \\
Training samples & 50{,}000 \\
\bottomrule
\end{tabular}
\end{table}

We optimize the regression objective with mean squared error (MSE) on the logarithmic targets using AdamW with a linear warmup schedule. Evaluation reports mean absolute error (MAE) on token-length prediction and standard classification metrics (accuracy, precision, recall, F1) for the long/short decision. We select the best checkpoint by validation accuracy. This design yields both fine-grained length estimates for capacity planning and robust binary decisions for practical scheduling and routing.

\begin{table}[!htbp]
\centering
\caption{Performance of the request classifier on testing datasets.}
\label{tab:req-metrics}
\setlength{\tabcolsep}{1pt}\begin{tabular}{lcccccc}
\toprule
 & Accuracy & F1 & Precision & Recall & \multicolumn{2}{c}{Regression} \\
\cmidrule(lr){6-7}
 & & & & & MAE (tokens) & \\
\midrule
Value & 0.8715 & 0.7909 & 0.7602 & 0.8243 & 110.01 \\
\bottomrule
\end{tabular}
\end{table}

\begin{figure*}[htb]
    \centering
\includegraphics[width=0.9\linewidth]{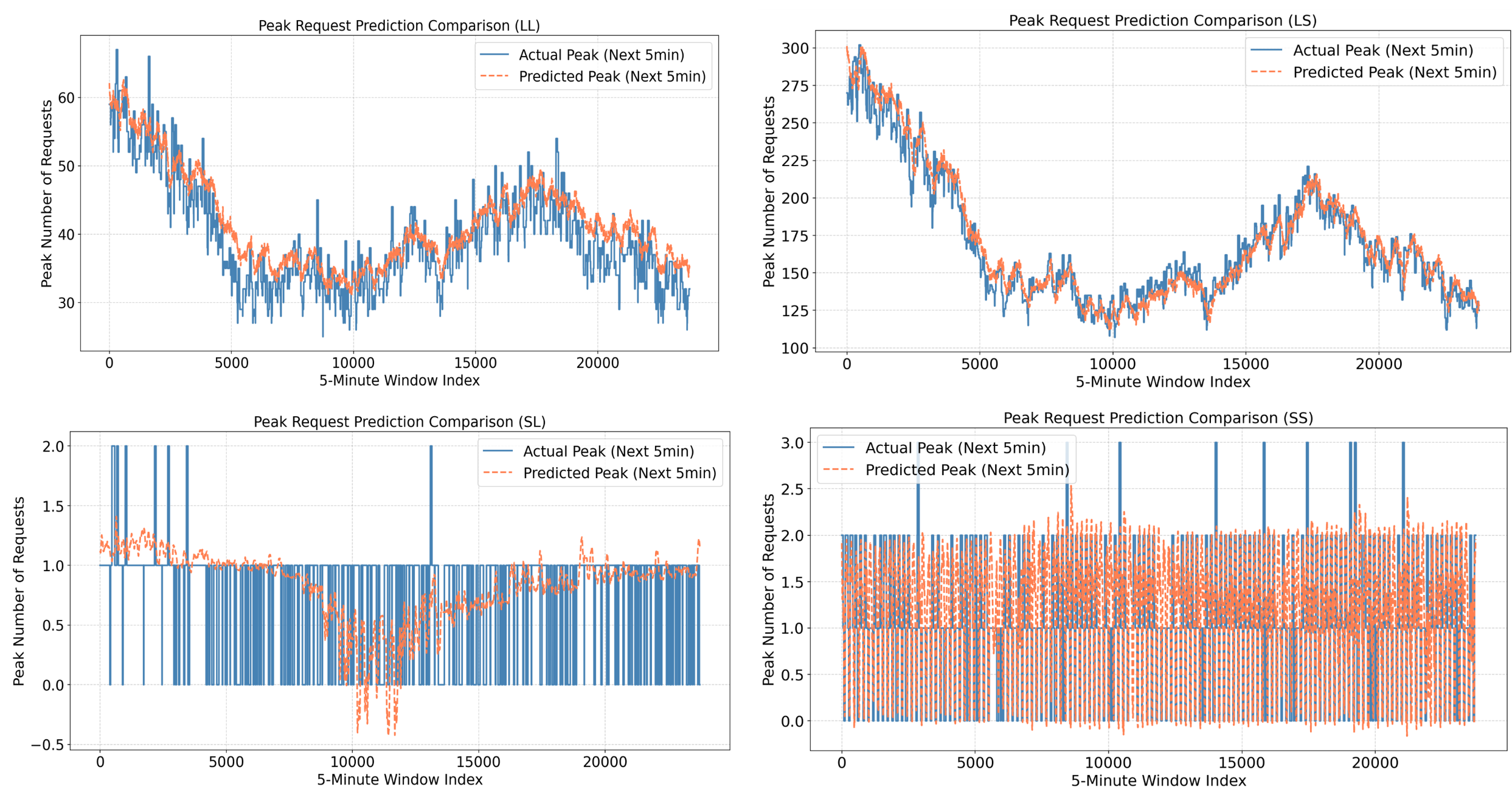}
    \caption{Comparison between LSTM-predicted and actual request traffic across different types.}
    \label{fig:lstm-prediction}
\end{figure*}
\subsection{Router and Worker} 
In our implementation, the \textit{Router} is responsible for receiving, classifying, and dispatching requests. Once a request is received from the client, the Router first extracts the complete prompt from the request body and computes the number of input tokens using the Llama~3 tokenizer. It then invokes an asynchronously wrapped BERT-based predictor to estimate the number of output tokens, where asynchronous execution prevents blocking of the event loop. Based on the input and predicted output token counts, the Router categorizes each request into one of four types: \emph{short-prompt--long-response}, \emph{short-prompt--short-response}, \emph{long-prompt--short-response}, and \emph{long-prompt--long-response}. After classification, the Router leverages pre-configured performance parameters associated with each type to estimate the end-to-end latency and determine the appropriate scheduling strategy.  

The Router operates in two distinct modes. In \textbf{Direct mode}, requests are dispatched strictly in order of arrival, effectively implementing a FCFS policy, without any further urgency-based prioritization. In \textbf{LLF mode}, by contrast, each request is assigned a task ID and placed into a laxity-based scheduling queue. The Router’s dispatcher continuously selects requests from the queue according to their urgency and forwards them to the appropriate worker. Regardless of the mode, worker selection follows a weighted round-robin strategy within the pool of available workers of the corresponding request type, and requests are forwarded to vLLM service instances through an OpenAI-compatible API. Upon task completion, the Router records metrics such as end-to-end latency, satisfaction of service-level objectives (SLOs), and the actual number of output tokens, which are later used for fairness and performance analysis.  

On the worker side, a vLLM service instance can register itself with the Router after initialization. Registration is performed through a management interface, where the worker provides its IP address, port, supported request type, and scheduling weight. The Router verifies the health of the worker before admitting it into the corresponding worker pool. The weight parameter directly affects the weighted round-robin policy, determining the proportion of requests assigned to the worker and thereby enabling resource-aware distribution of load across heterogeneous instances. Through this cooperative process, the Router and workers jointly establish a complete workflow that spans request reception, token-based classification, scheduling and dispatching, and result logging.  

\subsection{MIAD-based Frequency Scaling}

We implement the MIAD strategy to dynamically adjust GPU frequency for the \emph{vLLM} serving engine~\cite{kwon2023efficient}, allowing it to adapt to varying workload conditions.

To determine the optimal parameters for the MIAD controller, specifically the MI factor $M$ and the AD step $\delta$, we conducted a parameter tuning study. The experiment was initially conducted on a single NVIDIA A100 (40GB) GPU running the Llama3-8B model, and the frequency is kept the same a single across TP cluster. We simulated a 20-minute workload composed exclusively of short-prompt, short-answer (SS) requests.

We evaluated five configurations: MIAD (1.5, 10), MIAD (1.5, 100), MIAD (2.0, 10), MIAD (2.0, 100), and Fixed GPU frequency.
For brevity, each MIAD configuration is denoted by its MI factor and AD step (in MHz).
The Fixed baseline keeps the GPU frequency constant at its maximum value of 1410 MHz, which is the standard practice for LLM serving systems.

\begin{table*}[htbp]
\centering
\footnotesize 
\caption{MIAD Parameter Tuning Metrics}
\label{tab:miad_tuning_combined}

\begin{subtable}{\textwidth}
\centering
\caption{E2E and TTFT latency metrics}
\label{tab:e2e_ttft_metrics}
\begin{tabular*}{\textwidth}{@{\extracolsep{\fill}}l ccc ccc}
\toprule
Experiment & E2E p50 (ms) & E2E p90 (ms) & E2E p99 (ms) & TTFT p50 (ms) & TTFT p90 (ms) & TTFT p99 (ms) \\
\midrule
MIAD (1.5, 10) & 2866.98 & 5050.61 & 6008.65 & 64.26 & 127.43 & 173.39 \\
MIAD (1.5, 100) & 4151.74 & 5999.28 & 7290.99 & 93.64 & 169.15 & 195.64 \\
MIAD (2.0, 10) & 2497.30 & 4448.84 & 5640.99 & 53.97 & 105.68 & 154.71 \\
MIAD (2.0, 100) & 3424.21 & 5520.42 & 6510.17 & 77.48 & 159.10 & 197.39 \\
Fixed & \textbf{1520.67} & \textbf{1987.36} & \textbf{2414.20} & \textbf{33.76} & \textbf{40.47} & \textbf{44.52} \\
\bottomrule
\end{tabular*}
\end{subtable}

\vfill 

\begin{subtable}{\textwidth}
\centering
\caption{TBT, energy and violation metrics}
\label{tab:tbt_energy_violation}
\begin{tabular*}{\textwidth}{@{\extracolsep{\fill}}l ccc cc}
\toprule
Experiment & TBT p50 (ms) & TBT p90 (ms) & TBT p99 (ms) & Energy (Wh) & Violation Rate \\
\midrule
MIAD (1.5, 10) & 39.84 & 78.85 & 106.93 & 42.426 & 0.33\% \\
MIAD (1.5, 100) & 56.27 & 103.03 & 122.68 & \textbf{38.571} & 0.49\% \\
MIAD (2.0, 10) & 33.33 & 66.97 & 95.55 & 45.351 & 0.31\% \\
MIAD (2.0, 100) & 46.89 & 97.60 & 120.10 & 40.401 & 0.34\% \\
Fixed & \textbf{20.62} & \textbf{24.96} & \textbf{28.19} & 83.917 & \textbf{0.11\%} \\
\bottomrule
\end{tabular*}
\end{subtable}
\end{table*}

\begin{figure}[htbp]
    \centering
    \subfloat[MIAD (MI=1.5, AD=10)]{%
        \includegraphics[width=0.98\linewidth]{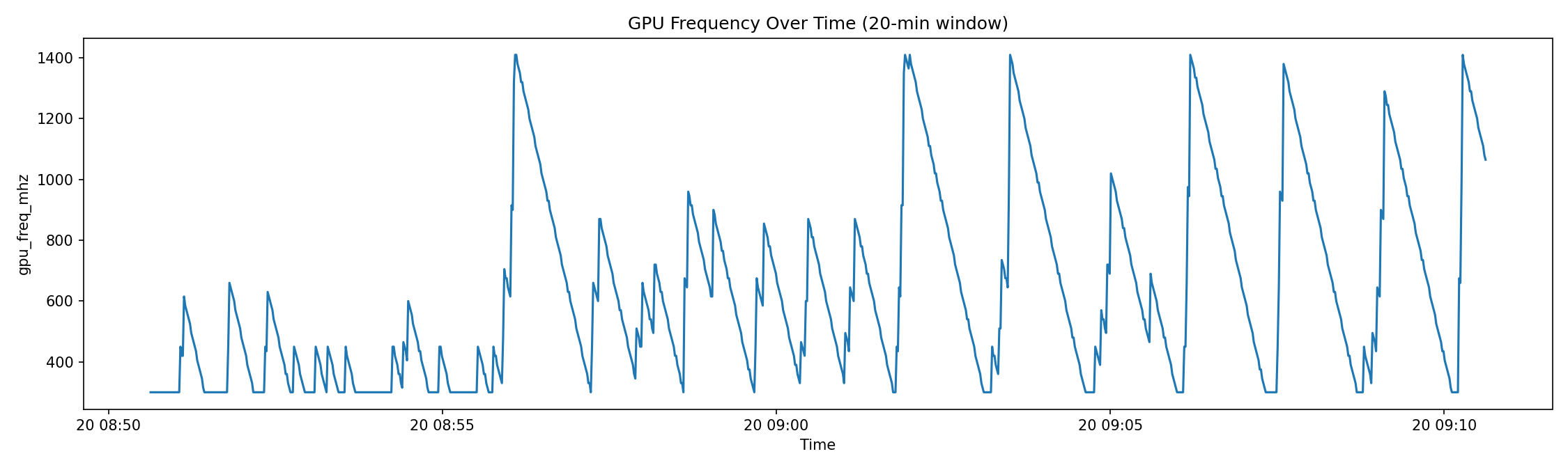}
        \label{fig:miad_15_10}
    }
    \vspace{1em}
    \subfloat[MIAD (MI=1.5, AD=100)]{%
        \includegraphics[width=0.98\linewidth]{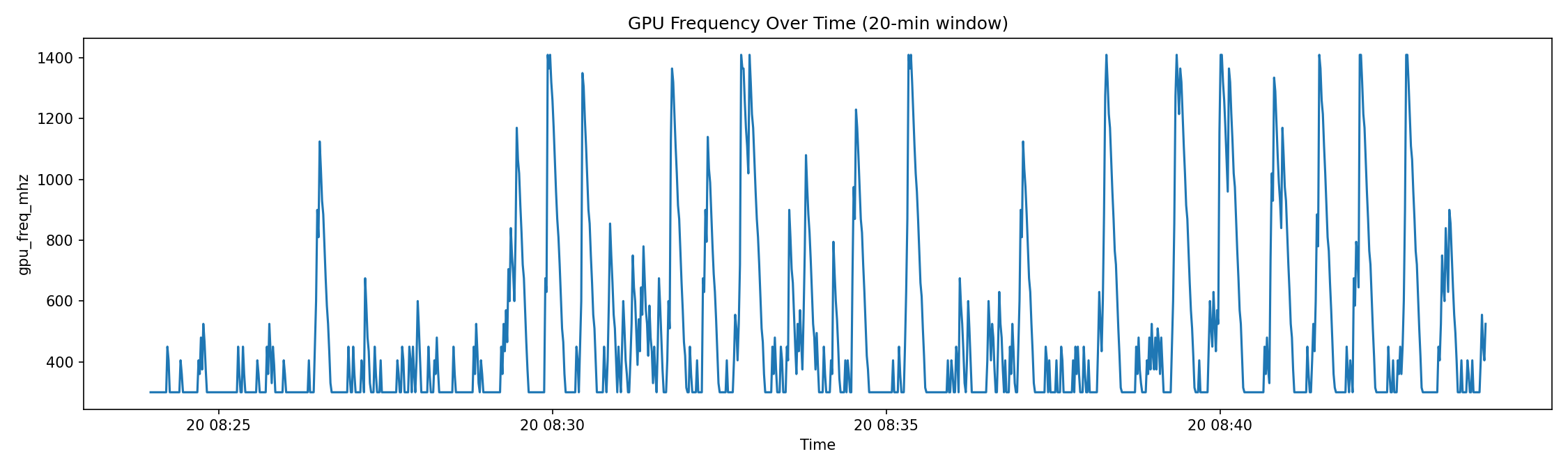}
        \label{fig:miad_15_100}
    }
    \vspace{1em} 
    \subfloat[MIAD (MI=2.0, AD=10)]{%
        \includegraphics[width=0.98\linewidth]{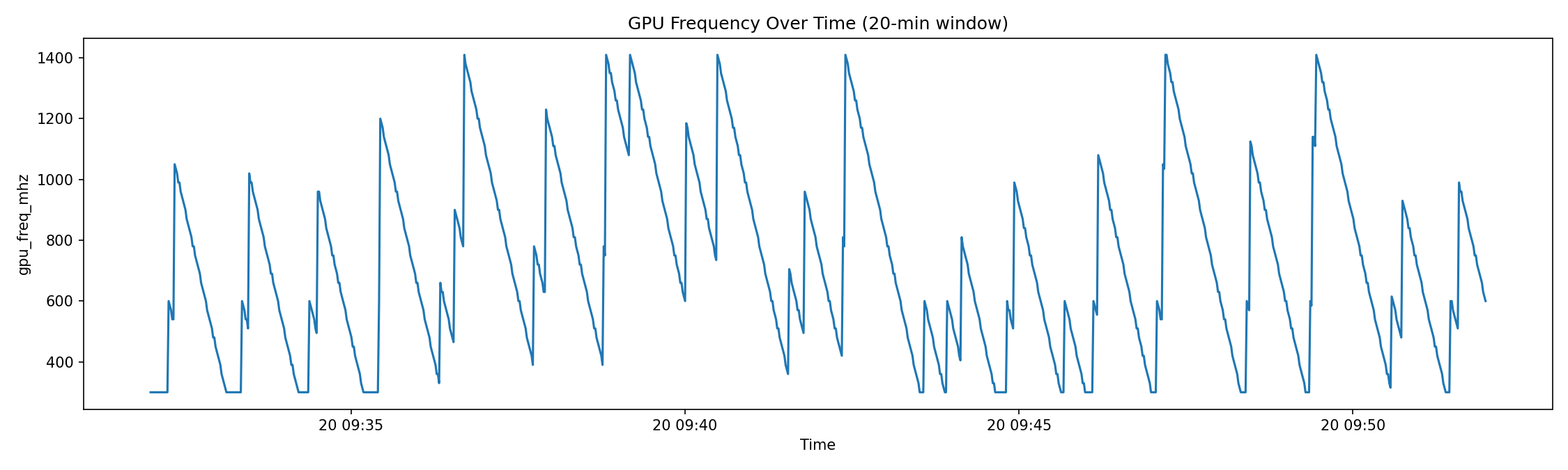}
        \label{fig:miad_2_10}
    }
    \vspace{1em} 
    \subfloat[MIAD (MI=2.0, AD=100)]{%
        \includegraphics[width=0.98\linewidth]{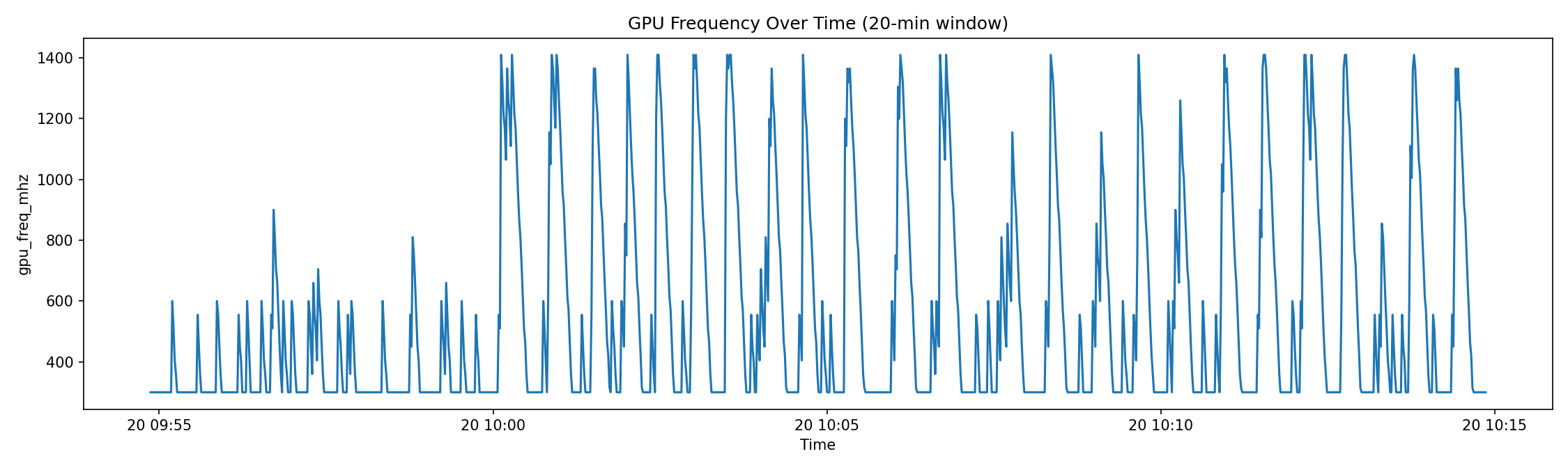}
        \label{fig:miad_2_100}
    }
    \caption{Frequency variation over 20 minutes for the four MIAD configurations under the SS workload.}
    \label{fig:miad_tuning_freq}
\end{figure}

The results are presented in Table \ref{tab:miad_tuning_combined}. The Fixed baseline, while offering the best latency (e.g., 20.62 ms TBT p50) and the lowest violation rate (0.11\%), incurred a prohibitively high energy cost (83.917 Wh). We therefore focused on identifying the best trade-off among the MIAD configurations.

Among these, MIAD (1.5, 100) achieved the lowest energy consumption (38.571 Wh), but at the expense of higher latencies and a higher violation rate (0.49\%). In contrast, MIAD (2.0, 10) delivered the best latency and lowest violation rate (0.31\%) within the MIAD group, though with the highest energy consumption (45.351 Wh).

Overall, MIAD (2.0, 100) provided the most balanced trade-off. Its energy consumption (40.401 Wh) was only slightly higher than the most efficient variant, while the violation rate (0.34\%) was considerably lower and the latency metrics remained moderate and acceptable. Based on the results of this parameter tuning study, we fix the MIAD parameters to (MI = 2.0, AD = 100) and integrate the controller into the full FREESH serving system. The following analysis evaluates its effectiveness in large-scale, heterogeneous serving environments.

As shown in Figure~\ref{fig:miad_freq_power_comparison}, the MIAD mechanism significantly reduces GPU frequency and power consumption compared to the \texttt{baseline}, while Figure~\ref{fig:miad_zoom_detail} provides a zoomed-in view that illustrates the fine-grained adjustment of GPU frequency in response to congestion events.

The specific latency-based thresholds used to trigger this mechanism (i.e., to define a "violation") are detailed by request type in Section D.2, Table~\ref{tab:slo_thresholds}. This parameter selection is aligned with the algorithmic formulation in Eq.~\eqref{eq:miad_alg} and is designed to balance SLO compliance with energy efficiency, avoiding excessive frequency oscillations.

It is worth noting that the MIAD strategy in FRESSH is closely related to \textmu-Serve~\cite{qiu2024power}. 
While FREESH and \textmu-Serve differ fundamentally in both the abstraction level and optimization objectives of frequency scaling. 
The core contribution of \textmu-Serve lies in identifying the heterogeneity of operator-level sensitivity to GPU frequency in LLM inference. 
Through offline profiling, \textmu-Serve co-designs operator partitioning and placement on homogeneous clusters, assigning frequency-insensitive (e.g., memory-bound) operators to specific GPUs. 
This enables the MIAD controller to substantially reduce power consumption without violating SLOs---a form of \textbf{micro-level (operator-level)} optimization. 

In contrast, FREESH addresses and adapts to a more comprehensive, marco-level, \emph{carbon-aware scheduling problem} across heterogeneous and geographically distributed GPU clusters, where operator-level analysis and deployment are impractical. 
Our MIAD mechanism therefore operates not on operator sensitivity but as a lightweight, \emph{request-level} feedback controller driven by latency-based SLO thresholds (as defined in Section~D.2). We also draw connections to resource allocation and utility maximization problem, thereby theoretically showing the global convergence to optimal frequency $f^*$ via MIAD.
The macro design also allows MIAD to be efficiently decoupled and also coordinated with other layers of the framework---namely, the slow-timescale pool-level carbon-aware routing and the real-time request-level LLF scheduling---jointly balancing carbon footprint, energy efficiency, and service fairness at scale.


\end{document}